\documentclass[english,prd,twocolumn,superscriptaddress,floatfix,nofootinbib,preprintnumbers,eqsecnum]{revtex4-1}
\usepackage{here}
\usepackage{graphicx}
\usepackage{amsmath,amsthm,amssymb}
\usepackage{bm}

\usepackage{amsfonts}
\usepackage{dcolumn}
\usepackage{hyperref}

\begin{document}
\newcommand{\newc}{\newcommand}

\newcommand{\ben}{\begin{eqnarray}}
\newcommand{\een}{\end{eqnarray}}
\newc{\be}{\begin{equation}}
\newc{\ee}{\end{equation}}
\newc{\ba}{\begin{eqnarray}}
\newc{\ea}{\end{eqnarray}}
\newc{\bea}{\begin{eqnarray*}}
\newc{\eea}{\end{eqnarray*}}
\newc{\D}{\partial}
\newc{\ie}{{\it i.e.} }
\newc{\eg}{{\it e.g.} }
\newc{\etc}{{\it etc.} }
\newc{\etal}{{\it et al.}}
\newcommand{\nn}{\nonumber}
\newc{\ra}{\rightarrow}
\newc{\lra}{\leftrightarrow}
\newc{\lsim}{\buildrel{<}\over{\sim}}
\newc{\gsim}{\buildrel{>}\over{\sim}}
\newc{\aP}{\alpha_{\rm P}}

\title{Cosmology in beyond-generalized Proca theories}

\author{
Shintaro Nakamura, 
Ryotaro Kase, and
Shinji Tsujikawa}

\affiliation{
Department of Physics, Faculty of Science, 
Tokyo University of Science, 1-3, Kagurazaka,
Shinjuku-ku, Tokyo 162-8601, Japan}

\date{\today}

\begin{abstract}
The beyond-generalized Proca theories are the extension of second-order 
massive vector-tensor theories (dubbed generalized Proca theories) with 
two transverse vector modes and one longitudinal scalar besides 
two tensor polarizations. 
Even with this extension, the propagating degrees of 
freedom remain unchanged on the isotropic cosmological background 
without an Ostrogradski instability.
We study the cosmology in beyond-generalized Proca theories by 
paying particular attention to the dynamics of late-time cosmic acceleration 
and resulting observational consequences.
We derive conditions for avoiding ghosts and  
instabilities of tensor, vector, and scalar perturbations and discuss viable parameter 
spaces in concrete models allowing the dark energy equation of state
smaller than $-1$. The propagation speeds of those perturbations
are subject to modifications beyond the domain of generalized Proca theories. 
There is a mixing between scalar and matter sound speeds, 
but such a mixing is suppressed 
during most of the cosmic expansion history without causing a new instability.
On the other hand, we find that derivative interactions arising in 
beyond-generalized Proca 
theories give rise to important modifications to the cosmic growth history.
The growth rate of matter perturbations can be compatible with 
the redshift-space distortion data due to the realization of gravitational 
interaction weaker than that in generalized Proca theories. 
Thus, it is possible to distinguish the dark energy model in beyond-generalized 
Proca theories from the counterpart in generalized Proca theories 
as well as from the $\Lambda$CDM model.
\end{abstract}


\maketitle

\section{Introduction}

After the first discovery of late-time cosmic acceleration \cite{SNIa}, 
the constantly accumulating observational data of supernovae Ia \cite{JLA}, 
cosmic microwave background (CMB) \cite{CMB}, and baryon acoustic oscillations \cite{BAO} have placed tighter bounds on the dark energy equation 
of state $w_{\rm DE}$. 
The cosmological constant $\Lambda$ (characterized by $w_{\rm DE}=-1$) is 
overall consistent with the observational data at background level, 
but the phantom equation of state ($w_{\rm DE}<-1$) is also allowed from 
the data \cite{Planck2015}. 
At the level of perturbations, the cosmic growth rate measurements of 
redshift-space distortions (RSD) \cite{Beu,Ledo,Eriksen} 
and cluster counts \cite{Vik} have shown tensions with the 
Planck CMB bound on $\sigma_8$ predicted by the 
$\Lambda$-cold-dark-matter ($\Lambda$CDM) 
model \cite{Planckdark}.

If the cosmological constant originates from the vacuum energy associated with 
particle physics, the vacuum energy usually acquires quantum corrections much larger 
than the observed dark energy scale \cite{Weinberg,Jerome}.
It is worth pursuing the possibility of realizing $w_{\rm DE}$ smaller than $-1$ 
without theoretical pathology while modifying gravitational interactions with matter 
to be consistent with the cosmic growth data. 
Modified gravitational theories can allow for 
the realization of such a possibility \cite{moreview1,moreview2}.

In the presence of a scalar field coupled to gravity, it is known that Horndeski 
theories \cite{Horndeski} are the most general scalar-tensor theories with second-order equations of motion. There are models of the late-time cosmic acceleration in the framework of 
Horndeski theories---like those based on $f(R)$ gravity \cite{fR}, Brans-Dicke 
theories \cite{Brans,Yoko}, and Galileons \cite{Galileon1,Galileon2,Galileon3,Galileon4}. 
These models can lead to $w_{\rm DE}$ smaller than $-1$ without having ghost and instability 
problems \cite{fR,Yoko,Galileon4}. In these models, the effective gravitational 
coupling $G_{\rm eff}$ of cosmological perturbations is usually larger than the Newton 
constant $G$ \cite{fRper,Yoko,Galima,DKT,Perenon}, so the growth rate of matter perturbations 
is enhanced compared to that in the $\Lambda$CDM model.

It is possible to perform a healthy extension of Horndeski theories in such a way that 
the number of propagating degrees of freedom (one scalar and two tensor modes) 
does not increase (see Ref.~\cite{Zuma} for an early work). 
Gleyzes-Langlois-Piazza-Vernizzi (GLPV) \cite{GLPV} 
expressed the Horndeski action in terms of scalar quantities arising in the 3+1 
decomposition of space-time \cite{building} and derived new derivative interactions 
without imposing two conditions Horndeski theories obey. 
In a nutshell, there are six free functions $A_2,~A_3,~ A_4,~ A_5,~ B_4,$ and $B_5$ in GLPV 
theories, whereas in Horndeski theories, the functions $B_4$ and $B_5$ are 
related to $A_4$ and $A_5$ respectively.
The beyond-Horndeski interactions of GLPV theories can give rise to several 
interesting effects such as the mixing of scalar and matter sound 
speeds \cite{Gergely,GLPV}, modified growth of subhorizon perturbations 
with additional time derivatives \cite{Koyama,weakgra}, and 
the appearance of solid-angle-deficit singularities \cite{conical}. 

For example, the covariantized Galileon 
model, the Lagrangian of which is derived by replacing partial derivatives of 
the Minkowski Galileon \cite{Galileon1} with covariant derivatives, 
belongs to a class of 
GLPV theories. This is different from the covariant Galileon model \cite{Galileon2} 
in which gravitational counterterms are added to keep the equations of motion 
up to second order. While the covariant Galileon is not excluded as 
a theoretically consistent dark energy model \cite{Galileon4}, the covariantized Galileon
is plagued by the problem of a negative scalar sound speed squared 
induced by the scalar-matter mixing \cite{Kase14}.
Thus, the extension outside the Horndeski domain generally leads to
nontrivial effects on the evolution of perturbations.

The scalar field is not the only possibility for driving the cosmic acceleration, 
but the vector field can be also the source for 
dark energy \cite{earlydark}. 
The massive vector field in Minkowski space-time (Proca theory) has 
one longitudinal scalar and two transverse vector modes due to 
the breaking of $U(1)$ gauge invariance.
If the massive vector field $A^{\mu}$ is coupled to gravity, it is possible to construct 
second-order generalized Proca (GP) theories by keeping three propagating 
degrees of freedom besides two tensor 
polarizations \cite{Heisenberg,Tasinato,Jimenez16} 
(see also Refs.~\cite{Horndeski2,Barrow,Jimenez13,TKK,Fleury,Hull,Li,Allys,Allys2,Minami,Cisterna,Amado,Emami}).

The existence of derivative interactions like those appearing for covariant 
vector Galileons in GP theories gives rise to a branch of background 
cosmological solutions where the temporal vector component $\phi$ 
depends on the Hubble expansion rate $H$ 
alone \cite{Tasinato,DeFelice16,Tomi}. 
For the covariant extended vector Galileon model
in which the Lagrangians contain general powers of 
$X=-A^{\mu}A_{\mu}/2$, there exists a de Sitter attractor 
preceded by the dark energy evolution with $w_{\rm DE}<-1$.
The effective gravitational coupling $G_{\rm eff}$ of cosmological 
perturbations is affected by the presence of intrinsic vector modes
in such a way that both $G_{\rm eff}<G$ and $G_{\rm eff}>G$
are possible \cite{Geff}. The screening mechanism of fifth forces 
in local regions of the Universe can be also at work in the presence 
of cubic and quartic derivative interactions \cite{scvector}.

Analogous to the extension of Horndeski theories to GLPV theories, 
it is possible to extend second-order GP theories to the domain of 
beyond-generalized Proca (BGP) theories \cite{HKT} 
(see also Ref.~\cite{Kimura}).
New BGP derivative couplings introduced in Ref.~\cite{HKT} constitute 
quartic and quintic scalar interactions as well as quintic and 
sixth-order vector interactions. Even in the presence of such interactions, 
there are no additional dangerous degrees of freedom associated 
with the Ostrogradski ghost on both isotropic 
and anisotropic cosmological backgrounds \cite{HKT,aniproca}. 
Moreover, unlike in GLPV theories, it was shown that solid-angle-deficit 
singularities do not generally arise in BGP theories due to the existence 
of the temporal vector component \cite{HKTconi}.

If we apply BGP theories to cosmology, it is not clear whether the new 
interactions mentioned above cause instabilities associated 
with the mixing of scalar and matter propagation speeds in concrete dark energy 
models. Moreover, it is of interest to study whether there are some 
distinct observational signatures of BGP theories as compared to 
GP theories and the $\Lambda$CDM model.
To address these issues, we study the cosmology based on 
the covariantized extended vector Galileon model in which partial derivatives of the 
extended vector Galileon in Minkowski space-time are replaced 
with covariant derivatives. We first discuss viable parameter spaces
consistent with no-ghost and stability conditions of tensor, 
vector, and scalar perturbations in the small-scale limit. 
We show that the mixing 
of sound speeds does not cause problems and that 
there are interesting observational signatures of weak gravity 
consistent with the recent RSD and CMB measurements. 

Our paper is organized as follows.
In Sec.\,\ref{beyondsec}, we review the action of BGP theories, 
and in Sec.\,\ref{backsec}, we discuss the background 
cosmology in the covariantized extended vector Galileon model. 
In Secs.~\ref{tensec} and \ref{vecsec}, we study no-ghost 
and stability conditions of tensor/vector perturbations 
and search for theoretically consistent parameter spaces. 
In Sec.\,\ref{scasec}, we present scalar perturbation 
equations of motion and study the mixing of sound speeds
for the covariantized extended vector Galileon model in detail.
In Sec.\,\ref{effsec}, we study the evolution of matter perturbations 
as well as gravitational potentials and show the possibility of 
observationally distinguishing dark energy models 
in BGP theories from those 
in GP theories and the $\Lambda$CDM model. 
We conclude in Sec.\,\ref{consec}.

\section{Beyond-generalized Proca theories}
\label{beyondsec}

We consider a massive vector field $A_{\mu}$ coupled to gravity with 
the field tensor $F_{\mu \nu}=\nabla_{\mu}A_{\nu}-\nabla_{\nu}A_{\mu}$, 
where $\nabla_{\mu}$ is a covariant derivative operator.
The mass term explicitly breaks a $U(1)$ gauge symmetry, 
so the longitudinal scalar mode arises in addition 
to two transverse vector polarizations. 
In GP theories with derivative couplings to 
gravity \cite{Heisenberg,Jimenez16}, 
the equations of motion for the vector field and the metric remain 
of second order.

It is possible to extend GP theories 
in such a way that the number of propagating degrees 
of freedom does not increase relative to those 
in GP theories (one scalar, two vectors, and
two tensors) \cite{HKT}.
The four-dimensional action of such extended 
theories (BGP theories) is given by 
\be
S=\int d^4 x \sqrt{-g} \left( \sum_{i=2}^{6} {\cal L}_i
+{\cal L}^{\rm N} \right)
+S_M (g_{\mu \nu}, \Psi_M)\,,
\label{action}
\ee
where $g$ is a determinant of the metric tensor 
$g_{\mu \nu}$, and 
\ba
{\cal L}_2 &=& G_2(X,F,Y)\,,
\label{L2}\\
{\cal L}_3 &=& G_3(X) \nabla_{\mu}A^{\mu}\,,
\label{L3}\\
{\cal L}_4 &=& 
G_4(X)R \nonumber \\
& &+
G_{4,X}(X) \left[ (\nabla_{\mu} A^{\mu})^2
-\nabla_{\rho}A_{\sigma}
\nabla^{\sigma}A^{\rho} \right]\,,\label{L4} \\
{\cal L}_5 &=& 
G_{5}(X) G_{\mu \nu} \nabla^{\mu} A^{\nu}
-\frac16 G_{5,X}(X) [ (\nabla_{\mu} A^{\mu})^3 \nonumber \\
&&-3\nabla_{\mu} A^{\mu}
\nabla_{\rho}A_{\sigma} \nabla^{\sigma}A^{\rho} 
+2\nabla_{\rho}A_{\sigma} \nabla^{\gamma}
A^{\rho} \nabla^{\sigma}A_{\gamma}] \nonumber \\
& &-g_5(X) \tilde{F}^{\alpha \mu}
{\tilde{F^{\beta}}}_{\mu} \nabla_{\alpha} A_{\beta}\,,
\label{L5}\\
{\cal L}_6 &=& G_6(X) L^{\mu \nu \alpha \beta} 
\nabla_{\mu}A_{\nu} \nabla_{\alpha}A_{\beta} \nonumber\\
&&+\frac12 G_{6,X}(X) \tilde{F}^{\alpha \beta} \tilde{F}^{\mu \nu} 
\nabla_{\alpha}A_{\mu} \nabla_{\beta}A_{\nu}\,,
\label{L6}
\ea
with 
\be
X=-\frac{A_{\mu} A^{\mu}}2 \,,\quad
F=-\frac{F_{\mu \nu} F^{\mu \nu}}4 \,,\quad
Y= A^{\mu}A^{\nu} {F_{\mu}}^{\alpha} 
F_{\nu \alpha}\,.
\label{Xdef}
\ee
While $G_2$ is a function of $X,F,Y$,   
the functions $G_{3,4,5,6}$ and $g_5$ depend on $X$ alone. 
For partial derivatives with respect to $X$, 
we use the notation $G_{i,X} \equiv \partial G_{i}/\partial X$. 
There are nonminimal derivative couplings of the vector 
field with the Ricci scalar $R$ and 
the Einstein tensor $G_{\mu \nu}$ in 
${\cal L}_4$ and ${\cal L}_5$, respectively.
In ${\cal L}_6$, there is also a derivative coupling with 
the double dual Riemann tensor defined by 
\be
L^{\mu \nu \alpha \beta}=\frac14 {\cal E}^{\mu \nu \rho \sigma} 
{\cal E}^{\alpha \beta \gamma \delta} R_{\rho \sigma \gamma \delta}\,,
\ee
where $R_{\rho \delta \gamma \delta}$ is the Riemann 
tensor and ${\cal E}^{\mu \nu \rho \sigma}$ is the Levi-Civit\`{a} tensor obeying 
the normalization ${\cal E}^{\mu \nu \rho \sigma}{\cal E}_{\mu \nu \rho \sigma}=-4!$.
For constant $G_6$, the coupling 
$G_6 L^{\mu \nu \alpha \beta} 
\nabla_{\mu}A_{\nu} \nabla_{\alpha}A_{\beta}$ is 
the only allowed $U(1)$ gauge-invariant interaction 
advocated by Horndeski \cite{Horndeski2}.
For $G_6$ depending on $X$, we need to introduce 
the second term in Eq.~(\ref{L6}) to keep the equations 
of motion up to second order (which is also the case 
for the second terms in ${\cal L}_4$ and ${\cal L}_5$). 
The quantity $\tilde{F}^{\mu \nu}$ is the dual strength 
tensor defined by 
\be
\tilde{F}^{\mu \nu}=\frac12 {\cal E}^{\mu \nu \alpha \beta}
F_{\alpha \beta}\,.
\ee
The terms $F$ and $Y$  in $G_2$ as well as
the terms containing the functions $g_5$ and $G_6$ 
correspond to intrinsic vector modes that vanish in the 
scalar limit $A_{\mu} \to \nabla_{\mu} \pi$.

The Lagrangian density ${\cal L}^{\rm N}$ in the action 
(\ref{action}) arises outside the domain of GP theories. 
The explicit form of ${\cal L}^{\rm N}$ 
is given by \cite{HKT}
\be
{\cal L}^{\rm N}=
{\cal L}_4^{\rm N}+{\cal L}_5^{\rm N}+
\tilde{{\cal L}_5^{\rm N}}+
{\cal L}_6^{\rm N}\,,
\ee
where
\ba
\hspace{-0.8cm}
& &{\cal L}_4^{\rm N}
=f_4 \hat{\delta}_{\alpha_1 \alpha_2 \alpha_3 \gamma_4}^{\beta_1 \beta_2\beta_3\gamma_4}
A^{\alpha_1}A_{\beta_1}
\nabla^{\alpha_2}A_{\beta_2} 
\nabla^{\alpha_3}A_{\beta_3}\,, \label{L4N}\\
\hspace{-0.8cm}
& &{\cal L}_5^{\rm N}
=
f_5 \hat{\delta}_{\alpha_1 \alpha_2 \alpha_3 \alpha_4}^{\beta_1 \beta_2\beta_3\beta_4}
A^{\alpha_1}A_{\beta_1} \nabla^{\alpha_2} 
A_{\beta_2} \nabla^{\alpha_3} A_{\beta_3}
\nabla^{\alpha_4} A_{\beta_4}\,,\label{L5N} \\
\hspace{-0.8cm}
& &\tilde{{\cal L}}_5^{\rm N}
=
\tilde{f}_{5}
\hat{\delta}_{\alpha_1 \alpha_2 \alpha_3 \alpha_4}^{\beta_1 \beta_2\beta_3\beta_4}
A^{\alpha_1}A_{\beta_1} \nabla^{\alpha_2} 
A^{\alpha_3} \nabla_{\beta_2} A_{\beta_3}
\nabla^{\alpha_4} A_{\beta_4}\,,
\label{L5Nd}\\
\hspace{-0.8cm}
&&{\cal L}_6^{\rm N}
=\tilde{f}_{6}
 \hat{\delta}_{\alpha_1 \alpha_2 \alpha_3 \alpha_4}^{\beta_1 \beta_2\beta_3\beta_4}
\nabla_{\beta_1} A_{\beta_2} \nabla^{\alpha_1}A^{\alpha_2}
\nabla_{\beta_3} A^{\alpha_3} \nabla_{\beta_4} A^{\alpha_4}\,,
\label{L6N}
\ea
with $\hat{\delta}_{\alpha_1 \alpha_2\gamma_3\gamma_4}^{\beta_1 \beta_2\gamma_3\gamma_4}={\cal E}_{\alpha_1 \alpha_2\gamma_3\gamma_4}
{\cal E}^{\beta_1 \beta_2\gamma_3\gamma_4}$.
The functions $f_{4},f_{5}$ and $\tilde{f}_{5},\tilde{f}_{6}$ 
depend on $X$ alone. The Lagrangian densities 
(\ref{L4N})--(\ref{L6N}) were
constructed in such a way that the relative coefficients 
between $G_i$ and $G_{i,X}$ (where $i=4,5,6$) appearing 
in Eqs.\,(\ref{L4})--(\ref{L6}) are detuned. 
Even with these new interactions, the propagating degrees 
of freedom for linear perturbations on the isotropic 
Friedmann-Lema\^{i}tre-Robertson-Walker (FLRW) 
background are the same as those in GP theories \cite{HKT}.  
On an anisotropic cosmological background, 
it was also shown in Ref.~\cite{aniproca} that 
there are no additional ghostly degrees of freedom 
associated with the Ostrogradski instability.

In Eq.~(\ref{action}), $S_M$ is the action of matter fields $\Psi_M$.
We assume that the matter fields are minimally coupled to gravity.
Since the vector field has a direct coupling to gravity, the matter sector feels 
the vector propagation through gravitational interactions.
In BGP theories, it is known that the Lagrangian 
density ${\cal L}^{\rm N}$ 
leads to a mixing between the scalar sound speed of 
the vector field and the matter sound speed \cite{HKT}. 
This is the important difference between GP theories and 
BGP theories, so we will estimate modifications of the sound 
speeds induced by ${\cal L}^{\rm N}$ in concrete dark energy models 
in Sec.\,\ref{scasec}.

Moreover, the effective gravitational coupling $G_{\rm eff}$ 
associated with the growth of matter perturbations should be 
also subject to change by new interactions of 
BGP theories. In particular, extra time derivatives can arise in the
perturbation equations of motion, so the quasistatic 
approximation used in GP theories for subhorizon 
modes \cite{Geff} may lose its validity. 
In Sec.\,\ref{effsec}, we will study how the evolution of matter perturbations 
and gravitational potentials is affected by the new interactions 
${\cal L}^{\rm N}$ in dark energy models 
within the framework of BGP theories.

Before entering the details of scalar perturbations, we will 
discuss the background cosmology and no-ghost and stability 
conditions of tensor and vector perturbations in subsequent 
sections to restrict the parameter space of dark energy models 
in BGP theories.

\section{Background cosmology}
\label{backsec}

\subsection{Background equations of motion}

On the flat FLRW space-time described by the line element 
$ds^2=-dt^2+a^2(t)\delta_{ij}dx^idx^j$, the background equations 
of motion were derived in Ref.~\cite{HKT}  in the presence of  
a matter perfect fluid with density $\rho_M$ and 
pressure $P_M$. The vector-field configuration compatible with the 
symmetry of the FLRW background contains the temporal 
component $\phi(t)$ alone, i.e., 
$A^{\mu}=(\phi(t),0,0,0)$. 
In Ref.~\cite{HKT}, it was shown that 
the background equations  depend only on four functions 
among ten free functions 
$G_{2,3,4,5,6},g_5,f_{4,5,},\tilde{f}_5,\tilde{f}_6$ 
appearing in the action (\ref{action}).
It is convenient to introduce the following combinations,
\ba
& &A_2=G_2\,,\qquad A_3=(2X)^{3/2} E_{3,X}\,, 
\nonumber\\
&& A_4=-G_4+2XG_{4,X}+4X^2f_4\,,\nonumber \\
& & A_5=-\sqrt{2}X^{3/2} \left( \frac13 G_{5,X}
-4Xf_5 \right)\,, \nonumber \\
 &&B_4=G_4\,, \qquad 
B_5=(2X)^{1/2}E_5\,, 
\label{ABre}
\ea
where $E_3(X)$ and $E_5(X)$ are auxiliary 
functions satisfying the relations
\be
G_3=E_3+2XE_{3,X}\,,\qquad 
G_{5,X}=\frac{E_5}{2X}+E_{5,X}\,.
\label{auxiliary}
\ee
Then, the gravitational equations of 
motion are given by \cite{HKT}
\ba
& &
A_2-6H^2 A_4-12H^3 A_5=\rho_M\,,
\label{back1}\\
& &
\dot{A}_3+4 (H \dot{A}_4+\dot{H}A_4) 
+6H ( H \dot{A}_5+2\dot{H}A_5 )=
\rho_M+P_M\,,\nonumber \\
\label{back2}
\ea
where $H=\dot{a}/a$ is the Hubble expansion rate 
and a dot represents a derivative with respect to $t$.
The temporal vector component obeys
\be
\phi \left( A_{2,X}+3HA_{3,X}+6H^2 A_{4,X}
+6H^3 A_{5,X} \right)=0\,,
\label{back3}
\ee
which can be also derived from 
Eqs.~(\ref{back1}) and (\ref{back2}).

{}From Eq.~(\ref{ABre}), it follows that 
\ba
A_4+B_4-2XB_{4,X}&=&4X^2f_4\,,\label{ABrelation0}\\
A_5+\frac13 XB_{5,X}&=&(2X)^{5/2}f_5\,. 
\label{ABrelation}
\ea
Since $f_4=0=f_5$ in GP theories, 
there are two particular relations 
$A_4+B_4-2XB_{4,X}=0$ and $A_5+XB_{5,X}/3=0$.
In BGP theories, the functions 
$B_4$ and $B_5$ are not directly related to $A_4$ 
and $A_5$ due to the existence of nonvanishing 
functions $f_4$ and $f_5$. 
The Lagrangians containing the functions 
$G_6, g_5, \tilde{f}_5, \tilde{f}_6$ do not contribute 
to the background equations of motion by reflecting the 
fact that they correspond to intrinsic vector modes.

{}From Eqs.~(\ref{back1})--(\ref{back3}), the background 
dynamics is determined by the four functions 
$A_{2,3,4,5}$, but it does not depend on 
$B_4$ and $B_5$. This means that BGP theories 
and GP theories with the same $A_{2,3,4,5}$ 
but with different $B_{4,5}$ cannot be distinguished 
from each other at the background level. 

\subsection{Covariant and covariantized 
extended vector Galileon models}

For concreteness, we consider the model given by 
the functions \cite{DeFelice16}
\be
G_2=b_2 X^{p_2}+F\,,\qquad 
G_3=b_3 X^{p_3}\,,
\label{G23}
\ee
and 
\be
G_4=\frac{M_{\rm pl}^2}{2}+b_4X^{p_4}\,,
\qquad 
G_5=b_5X^{p_5}\,,
\label{modela}
\ee
where $b_{2,3,4,5}$ and $p_{2,3,4,5}$ are constants, 
and $M_{\rm pl}=(8\pi G)^{-1}$ is the reduced Planck mass. 
{}From Eq.~(\ref{auxiliary}), we can choose 
the auxiliary functions $E_3$ and $E_5$ in 
the forms $E_3=b_3X^{p_3}/(1+2p_3)$ and 
$E_5=2b_5p_5X^{p_5}/(1+2p_5)$, respectively. 
Since $f_4=0=f_5$ in GP theories, 
the functions in Eq.~(\ref{ABre}) yield
\ba
\hspace{-0.5cm}
& &
A_2=b_2X^{p_2}+F\,,\qquad 
A_3=\frac{2\sqrt{2}b_3 p_3}{1+2p_3}X^{p_3+1/2}\,, 
\nonumber \\
\hspace{-0.5cm}
& &
A_4=-\frac{M_{\rm pl}^2}{2}+b_4(2p_4-1)X^{p_4}\,,
\nonumber \\
\hspace{-0.5cm}
& &
A_5=-\frac{\sqrt{2}}{3}b_5p_5 X^{p_5+1/2}\,,
\label{extendedco}
\ea
and 
\be
B_4=\frac{M_{\rm pl}^2}{2}+b_4X^{p_4}\,,
\qquad
B_5=\frac{2\sqrt{2}b_5p_5}{1+2p_5}X^{p_5+1/2}.
\label{extendedco2}
\ee
The covariant vector Galileon \cite{Heisenberg} 
corresponds to the powers 
$p_2=1, p_3=1, p_4=2, p_5=2$. 
The model given by more general functions (\ref{G23}) and 
(\ref{modela}) together with couplings
$g_5$ and $G_6$ is dubbed the covariant extended vector 
Galileon (covariant EVG). 
In this case, the functions $B_4$ and $B_5$ 
obey the relations $A_4+B_4-2XB_{4,X}=0$ and 
$A_5+XB_{5,X}/3=0$ to keep the equations of 
motion up to second order. 

In BGP theories, there are no particular constraints 
between $A_4, B_4, A_5,$ and $B_5$. 
Let us consider theories in which 
the functions $A_{2,3,4,5}$ are the same as those 
in Eq.~(\ref{extendedco}) but with the functions 
$B_4$ and $B_5$ given by 
\be
B_4=\frac{M_{\rm pl}^2}{2}\,,\qquad 
B_5=0\,.
\label{extendedco3}
\ee
In this case we have 
\be
G_4=\frac{M_{\rm pl}^2}{2}\,,\qquad 
G_5=0\,,
\label{modelb}
\ee
which are different from the functions $G_4$ and $G_5$ 
in Eq.~(\ref{modela}). 
Note that $G_5$ can be a nonvanishing constant, 
but we have set $G_5=0$ without loss of generality 
since the constant $G_5$ 
does not contribute to the dynamical equations of motion 
(due to the property $\nabla^{\mu}G_{\mu \nu}=0$).
The functions $f_4$ and $f_5$, which characterize the 
deviation from GP theories, are given, respectively, by 
\be
f_4=\frac{1}{4} b_4(2p_4-1)X^{p_4-2}\,,\qquad
f_5=-\frac{1}{12} b_5p_5 X^{p_5-2}\,.
\label{f45}
\ee
The Lagrangians of BGP theories we are considering now 
contain the interactions ${\cal L}_4^{\rm N}$ and 
${\cal L}_5^{\rm N}$ besides the 
Einstein-Hilbert term $(M_{\rm pl}^2/2)R$.
These new terms correspond to 
those derived by replacing partial derivatives 
with covariant derivatives for the EVG model
in Minkowski space-time (analogous to the covariantized 
Galileon model discussed in Ref.~\cite{Kase14}). 
The model given by  the functions (\ref{G23}), (\ref{modelb}), and 
(\ref{f45}) together with the other couplings 
$g_5, G_6, \tilde{f}_5, \tilde{f}_6$
is dubbed the covariantized EVG. 

The background cosmological dynamics in the 
covariantized EVG model is exactly the same as 
that in the covariant EVG model. 
Since the dynamics in the latter was studied 
in Ref.~\cite{DeFelice16}, we briefly 
summarize the main results. 
We consider the powers $p_{3,4,5}$ satisfying 
\ba
& &
p_3=\frac{1}{2} \left( p+2p_2-1 \right)\,,\qquad 
p_4=p+p_2\,,\nonumber \\
& &
p_5=\frac{1}{2} \left( 3p+2p_2-1 \right)\,,
\label{p345}
\ea
where $p$ is a positive constant. Then, the 
nonvanishing $\phi$ branch of Eq.~(\ref{back3}) 
gives rise to the following solution:
\be
\phi \propto H^{-1/p}\,.
\ee
Since $\phi$ grows with the decrease of $H$, 
the energy density of the temporal vector component 
works as dark energy at late cosmological epochs. 
In fact, there exist de Sitter solutions characterized 
by constant $\phi$ and $H$.

For the matter action $S_M$, we take into account 
the perfect fluids of nonrelativistic matter (density 
$\rho_m$ and pressure $P_m=0$) and 
radiation (density $\rho_r$ and pressure $P_r=\rho_r/3$). 
We define the corresponding density parameters 
$\Omega_m=\rho_m/(3M_{\rm pl}^2H^2)$ and 
$\Omega_r=\rho_r/(3M_{\rm pl}^2H^2)$ as 
well as the dark energy density parameter 
\be
\Omega_{\rm DE}=
\frac{\gamma}{p+p_2}y\,,
\label{Omede}
\ee
where 
\ba
\hspace{-0.5cm}
y 
&=& \frac{b_2\phi^{2p_2}}
{3M_{\rm pl}^2H^2 2^{p_2}}\,,
\label{ydef}\\
\hspace{-0.5cm}
\gamma
&=&6p_2 (2p+2p_2-1)\beta_4-(p+p_2)
(1+4p_2 \beta_5)\,,\\
\hspace{-0.5cm}
\beta_i
&=& \frac{p_ib_i(\phi^p H)^{i-2}}{2^{p_i-p_2}p_2b_2}\,,
\label{betai}
\ea
with $i=3,4,5$.
{}From Eq.~(\ref{back3}), we have the following relation 
for the branch $\phi \neq 0$:
\be
1+3\beta_3+6\left( 2p+2p_2-1 \right)\beta_4
-\left( 3p+2p_2 \right) \beta_5=0\,.
\ee

The background equations (\ref{back1}) and (\ref{back2}) 
can be expressed as
\ba
3M_{\rm pl}^2H^2&=&\rho_{\rm DE}+\rho_m+\rho_r\,,
\label{back1d}\\
-2M_{\rm pl}^2\dot{H}&=&\rho_{\rm DE}+P_{\rm DE}
+\rho_m+\frac{4}{3}\rho_r\,,
\label{back2d}
\ea
where $\rho_{\rm DE}$ and 
$P_{\rm DE}$ correspond to the density and the 
pressure associated with the vector field, respectively.
Introducing the ratio $s=p_2/p$, the dynamical equations 
of motion can be expressed in the autonomous 
forms \cite{DeFelice16}
\ba
\Omega_{\rm DE}'
&=& \frac{(1+s)\Omega_{\rm DE}(3+\Omega_r-3\Omega_{\rm DE})}{1+s\Omega_{\rm DE}}\,,
\label{auto1}\\
\Omega_{r}'
&=& -\frac{\Omega_r [1-\Omega_r+(3+4s)\Omega_{\rm DE}]}{1+s\Omega_{\rm DE}}\,,
\label{auto2}
\ea
where a prime represents a derivatives with respect to 
$N=\ln a$. The matter density parameter is known from 
the relation $\Omega_{m}=1-\Omega_{\rm DE}-\Omega_r$. 
The dark energy equation of state, which is defined by 
$w_{\rm DE}=P_{\rm DE}/\rho_{\rm DE}$, reads
\be
w_{\rm DE}=-\frac{3(1+s)+s\Omega_r}
{3(1+s\Omega_{\rm DE})}\,.
\label{wde}
\ee

For the dynamical system (\ref{auto1})--(\ref{auto2}), 
there exist the three fixed points: 
(a) radiation: $(\Omega_{\rm DE}, \Omega_r)=(0,1)$, 
(b) matter: $(\Omega_{\rm DE}, \Omega_r)=(0,0)$, and 
(c) de Sitter: $(\Omega_{\rm DE}, \Omega_r)=(1,0)$. 
During the cosmological sequence of the fixed points 
(a) $\to$ (b) $\to$ (c), the dark energy equation of state evolves as 
(a) $w_{\rm DE}=-1-4s/3$ $\to$
(b) $w_{\rm DE}=-1-s$ $\to$
(c) $w_{\rm DE}=-1$. 
{}From Eqs.~(\ref{auto1})--(\ref{auto2}), there is 
the relation $\Omega_{\rm DE}'/\Omega_{\rm DE}=(1+s)
(\Omega_{r}'/\Omega_{r}+4)$, which is integrated to give
  $\Omega_{\rm DE}
/\Omega_r^{1+s} \propto a^{4(1+s)}$. 
On using this solution, the evolution of $\Omega_{\rm DE}$ 
and $\phi$ during the radiation and the matter eras is given by 
\be
\Omega_{\rm DE} \propto t^{2(1+s)}\,,\qquad 
\phi \propto t^{1/p}\,.
\ee
For $s>-1$, the dark energy density parameter grows 
in time. Since the fixed point (c) is always 
stable \cite{DeFelice16}, the solutions finally approach 
the de Sitter attractor to give rise to the late-time 
cosmic acceleration. 

The covariant and the covariantized EVG models can be distinguished from 
each other at the level of linear cosmological perturbations. 
Since the perturbations can be decomposed into tensor, 
vector, and scalar modes, we will separately study the behavior of 
each mode in subsequent sections. 

\section{Tensor perturbations}
\label{tensec}

\subsection{Stability conditions}

We begin with tensor perturbations $h_{ij}$ given by 
the line element 
\be
ds_T^2=-dt^2+a^2(t) \left( \delta_{ij}+h_{ij} 
\right) dx^i dx^j\,.
\ee
Due to the transverse and traceless conditions 
$\partial^i h_{ij}=0$ and ${h_i}^i=0$, there are 
two polarization modes $h_{+}$ and $h_{\times}$ for 
$h_{ij}$. In terms of the unit vectors $e_{ij}^{+}$ 
and $e_{ij}^{\times}$ satisfying the normalizations
$e_{ij}^{+}({\bm k}) e_{ij}^{+}(-{\bm k})^*=1$,
$e_{ij}^{\times}({\bm k}) e_{ij}^{\times}(-{\bm k})^*=1$,
and $e_{ij}^{+}({\bm k}) e_{ij}^{\times}(-{\bm k})^*=0$ 
in Fourier space with wave number ${\bm k}$, 
we can express $h_{ij}$ in the form 
$h_{ij}=h_{+}e_{ij}^{+}+h_{\times}e_{ij}^{\times}$.
The second-order action of tensor perturbations for the theory (\ref{action}) is given by \cite{HKT}
\be
S_T^{(2)}=\sum_{\lambda={+},{\times}}\int dt\,d^3x\,
a^3\,\frac{q_T}{8}  \left[\dot{h}_\lambda^2
-\frac{c_T^2}{a^2}(\partial h_\lambda)^2\right]\,,
\label{ST}
\ee
where
\ba
q_T &=& -2 \left( A_4+3HA_5 \right)\,,
\label{qT}\\
c_T^2 &=& \frac{2B_4+\dot{B}_{5}}{q_T}\,.
\label{cT}
\ea
The conditions for avoiding ghosts and Laplacian instabilities 
correspond to $q_T>0$ and $c_T^2>0$, respectively. 
The ghost condition is determined by the quantities $A_4$ and $A_5$ 
appearing in the background equations of motion. 
Since $c_T^2$ contains $B_4$ and $B_5$, the theories 
with same values of $A_4, A_5$ but with different values 
of  $B_4,B_5$ can be distinguished from the tensor propagation speed. 
The intrinsic vector modes (associated with the functions 
$g_5,G_6,\tilde{f}_5,\tilde{f}_6$) affect neither $q_T$ nor $c_T^2$.

We write the function $A_4$ in the form
\be
A_4(X)=-\frac{M_{\rm pl}^2}{2}+\tilde{A}_4(X)\,,
\ee
where $\tilde{A}_4(X)$ is a function of $X$.
For covariant and covariantized EVG models, 
$\tilde{A}_4(X)=b_4(2p_4-1)X^{p_4}$.
{}From the background equations (\ref{back1})--(\ref{back2}),
the density $\rho_{\rm DE}$ and the pressure $P_{\rm DE}$ 
in Eqs.~(\ref{back1d}) and (\ref{back2d}) are given, 
respectively, by 
\ba
\rho_{\rm DE} 
&=& -A_2+6H^2 \tilde{A}_4+12H^3 A_5\,,\\
P_{\rm DE}
&=& A_2-\dot{A}_3-2(3H^2+2\dot{H})\tilde{A}_4 
-4H \dot{\tilde{A}}_4 \nonumber \\
& &-12H (H^2+\dot{H})A_5-6H^2 \dot{A}_5\,.
\ea
We are now considering the case in which $\rho_{\rm DE}$ and $P_{\rm DE}$ 
are responsible for the late-time cosmic acceleration. 
During the radiation and matter eras, both 
$\rho_{\rm DE}$ and $P_{\rm DE}$ are suppressed 
relative to the background density 
$\rho_M \approx M_{\rm pl}^2H^2$, so
the conditions
\be
\{ |A_2|, H|A_3|, H^2 |\tilde{A}_4|, H^3 |A_5| \} 
\ll M_{\rm pl}^2 H^2
\label{Acon}
\ee
are satisfied. Under these conditions, the quantity 
$q_T=M_{\rm pl}^2-2\tilde{A}_4-6HA_5$ is 
approximately given by 
\be
(q_T)_{\rm early} \simeq M_{\rm pl}^2\,,
\label{qtap}
\ee
which means that the tensor ghost is absent in 
the early cosmological epoch. 
In the late Universe, there are contributions 
from the terms $\tilde{A}_4$ and $A_5$ to $q_T$, 
but as long as the condition 
\be
2\tilde{A}_4+6H A_5<M_{\rm pl}^2
\ee
is satisfied, there is no tensor ghost.
 
For the estimation of $c_T^2$, we express  
$B_4$ in the form 
\be
B_4(X)=\frac{M_{\rm pl}^2}{2}+\tilde{B}_4(X)\,,
\ee
where $\tilde{B}_4(X)$ is a function of $X$. 
In the early cosmological epoch, the functions $A_4$ and 
$A_5$ should satisfy Eq.~(\ref{Acon}) to realize 
the consistent background dynamics, in which regime 
Eq.~(\ref{cT}) reduces to 
\be
(c_T^2)_{\rm early} 
\simeq 1+\frac{2\tilde{B}_4+\dot{B}_5}
{M_{\rm pl}^2}\,.
\ee 

In GP theories, the functions $\tilde{B}_4$ 
and $B_5$ are subject to the 
constraints $\tilde{A}_4+\tilde{B}_4-2X\tilde{B}_{4,X}=0$ 
and $3A_5+XB_{5,X}=0$, so they also satisfy the 
conditions similar to those of $A_4$ and $A_5$, i.e., 
$|\tilde{B}_4| \ll M_{\rm pl}^2$ and 
$H |B_5| \ll M_{\rm pl}^2$ for power-law functions
of $\tilde{B}_4$ and $B_5$. Provided that $|\dot{B}_5|$ is 
at most of the order of $H|B_5|$, we have that 
$c_T^2 \simeq 1$ in the early 
cosmological epoch in GP theories.

In BGP theories， the functions $\tilde{B}_4$ and $B_5$ are 
independent of $\tilde{A}_4$ and $A_5$, respectively, so $c_T^2$ 
is not necessarily close to 1 at high redshifts.
If the quantities $|\tilde{B}_4|$ and 
$H |B_5|$ are not much smaller than $M_{\rm pl}^2$ 
in the early cosmological epoch, 
the deviation of $c_T^2$ from 1 
is significant at low redshifts. 
This leads to either the Laplacian instability ($c_T^2<0$)
or the highly superluminal propagation speed squared 
($c_T^2 \gg 1$) being possibly incompatible with 
observational bounds of $c_T^2$. 
Then, it is safe to consider the situation in which 
the two conditions 
\be
|\tilde{B}_4| \ll M_{\rm pl}^2\,,\qquad 
H |B_5| \ll M_{\rm pl}^2
\label{Bcon}
\ee
are satisfied. In this case, $c_T^2 \simeq 1$ 
at high redshifts. 

On the de Sitter solution, we have $\dot{B}_5=0$, so 
Eq.~(\ref{cT}) reduces to 
\be
(c_T^2)_{\rm dS}
=\left( 1+\frac{2\tilde{B}_4}{M_{\rm pl}^2} \right)
\left( 1-\frac{2\tilde{A}_4}{M_{\rm pl}^2}
-\frac{6HA_5}{M_{\rm pl}^2} \right)^{-1}\,.
\label{ctds}
\ee
Since the inequality (\ref{Acon}) does not generally 
hold on the de Sitter solution, there is the deviation 
of $(c_T^2)_{\rm dS}$ from 1. Moreover, we also have 
the contribution to $(c_T^2)_{\rm dS}$ from the term 
$\tilde{B}_4$, which is different between GP theories 
and BGP theories.

\subsection{Covariant and covariantized EVG models}

For concreteness, we consider the covariant and covariantized 
EVG models in which the functions $A_{2,3,4,5}$ are 
given by Eq.~(\ref{extendedco}) with the powers (\ref{p345}). 
The difference between the two models arises from 
the functions $B_4$ and $B_5$.
Since the parameter (\ref{ydef})
satisfies the relation $y=(p+p_2)\Omega_{\rm DE}/\gamma$, 
the quantity (\ref{qT}) reduces to
\be
q_T=M_{\rm pl}^2 \left[ 1-\frac{6p_2}{\gamma}
\left\{ (2p+2p_2-1)\beta_4-(p+p_2)\beta_5 \right\}
\Omega_{\rm DE} \right]\,.
\label{qtex}
\ee
Since $\Omega_{\rm DE} \to 0$ in the asymptotic past, 
we recover the property (\ref{qtap}). 
On the de Sitter solution ($\Omega_{\rm DE}=1$),
the no-ghost condition ($q_T>0$) is satisfied for $|\beta_4|$ and 
$|\beta_5|$ much smaller than 1.

In the covariant EVG model, 
the background dynamics restricts $c_T^2$ to be close to 1 
in the early cosmological 
epoch. In the covariantized EVG model 
($\tilde{B}_4=0, B_5=0$), the conditions (\ref{Bcon}) are 
automatically satisfied, so $c_T^2 \simeq 1$ 
at high redshifts. 

On the de Sitter solution ($\Omega_{\rm DE}=1$), 
the tensor propagation speed squared (\ref{ctds}) in
the covariant EVG model ($\tilde{B}_4=b_4X^{p_4}$) reads
\be
(c_T^2)_{\rm dS 1}
=1-\frac{6p_2(2\beta_4-\beta_5)}{1-2p_2\beta_5}\,.
\label{ct1}
\ee
On the other hand, the covariantized EVG model corresponds to 
$\tilde{B}_4=0$, so Eq.~(\ref{ctds}) yields
\be
(c_T^2)_{\rm dS 2}
=1-\frac{6p_2[(2p+2p_2-1)\beta_4-(p+p_2)\beta_5]}
{(p+p_2)(1-2p_2 \beta_5)}\,.
\label{ct2}
\ee
For larger $|\beta_4|$ and $|\beta_5|$, both 
$(c_T^2)_{\rm dS 1}$ and $(c_T^2)_{\rm dS 2}$ tend 
to be away from 1. However, as long as the conditions 
\be
|\beta_4| \ll 1\,,\qquad |\beta_5| \ll 1
\label{beta45}
\ee
are satisfied, they do not deviate much from 1.
{}From Eqs.~(\ref{ct1}) and (\ref{ct2}), 
we have $(c_T^2)_{\rm dS 2}-(c_T^2)_{\rm dS 1}
=6p_2\beta_4/[(p+p_2)(1-2p_2\beta_5)]$, the 
difference of which gets more significant for larger $|\beta_4|$. 
In summary,  under the conditions (\ref{beta45}), 
there are neither ghosts nor Laplacian 
instabilities in both covariant and covariantized 
EVG models.

{}From the CMB observations, the tensor propagation 
speed squared is constrained to be $c_T^2=1.30 \pm 0.79$ 
at 95\,\%\,confidence level by assuming that $c_T^2$ 
is constant \cite{Raveri}. 
{}From the gravitational Cherenkov radiation, there exists 
the tight bound $1-c_T<2 \times 10^{-15}$ for the 
subluminal propagation\footnote{The superluminal case 
is not subject to the Cherenkov-radiation constraint. 
The superluminal propagation does not necessarily 
imply the violation of causality, since, in many theories 
including k-essence \cite{Babi} and Galileons \cite{GSami}, 
the appearance of closed causal curves can be avoided.} \cite{Moore}, 
but the corresponding energy, $\sim 10^{10}$~GeV, 
is much higher than any reasonable cutoff
associated with the late-time cosmic acceleration. 
{}From binary pulsars timing data, the deviation of $c_T$ 
from 1 is constrained to the level of $10^{-2}$ \cite{Fedo}.
Under the conditions (\ref{beta45}), $c_T^2$ is very close 
to 1 during the cosmic expansion history, so the above-mentioned 
observational bound of $c_T^2$ can be satisfied.

\section{Vector perturbations}
\label{vecsec}

\subsection{Stability conditions}

The vector perturbation arises from the spatial component 
$A^i$ of the vector field. 
We express the intrinsic vector mode $E_j$ in $A^i$, as 
$(A^i)_V=E_j \delta^{ij}/a^2(t)$, where $E_j$ satisfies 
the transverse condition $\partial^j E_j=0$. 
We also consider the metric perturbation $V_i$ 
described by the line element 
\be
ds_V^2=-dt^2+2V_i dt dx^i+a^2(t)\delta_{ij} dx^idx^j\,,
\ee
where we have chosen the flat gauge.
The vector perturbation also obeys the transverse 
condition $\partial^i V_i=0$. The combination 
\be
Z_i=E_i+\phi(t)V_i
\ee
corresponds to a dynamical degree of freedom with 
two transverse polarizations. The matter perfect fluid 
can be accommodated by the Schutz-Sorkin 
action \cite{Sorkin}, which does not propagate 
a new degree of freedom in the vector sector \cite{HKT}.

We choose the direction of the momentum ${\bm k}$ 
along the $z$ direction and consider the vector field 
in the form $Z_i=(Z_1(z), Z_2(z), 0)$.
Expanding the action (\ref{action}) up to second order 
in vector perturbations and taking the small-scale limit, 
the resulting second-order action for the two dynamical 
fields $Z_i$ reads \cite{HKT}
\be
S_V^{(2)} \simeq \int dt d^3 x \sum_{i=1}^{2} 
\frac{aq_V}{2} \left[ \dot{Z}_i^2 
-\frac{c_V^2}{a^2} \left( \partial Z_i \right)^2 
\right]\,,
\ee
where 
\ba
q_V 
&=& G_{2,F}+2G_{2,Y}\phi^2-4g_5H \phi+2G_6 H^2 
\nonumber \\
& &+2G_{6,X}H^2 \phi^2+4\tilde{f}_6 H^2 \phi^2\,,
\label{qV} \\
c_V^2
&=&
1+\frac{2(A_4+B_4+3HA_5)^2}{\phi^2 q_T q_V}
\nonumber \\
& &+\frac{2(G_6\dot{H}-G_{2,Y}\phi^2)
-2(H \phi-\dot{\phi})(G_{6,X}H \phi-g_5)}{q_V}
\nonumber \\
& &
-\frac{2}{q_V} \left[ \tilde{f}_5H \phi^3 
+2\tilde{f}_6H \phi (H \phi-\dot{\phi}) \right]\,.
\label{cV}
\ea

The functions $F$ and $Y$ in $G_2$ as well as the 
functions $g_5, G_6,$ and $\tilde{f}_6$ affect the quantity 
$q_V$ (which characterizes the vector no-ghost condition). 
Besides these intrinsic vector modes, the function 
$\tilde{f}_5$ also leads to a modification to the vector 
propagation speed $c_V$. 
The difference of $c_V^2$ between GP
theories and BGP theories arises 
through the functions $B_4$ (or $f_4$) and 
$\tilde{f}_5, \tilde{f}_6$.

To understand the effect of the terms beyond the domain 
of GP theories, we consider the theories 
of nonvanishing functions $\tilde{f}_5, \tilde{f}_6$, 
and 
\be
G_2=F+g_2(X)\,,\qquad 
g_5=0\,,\qquad G_6=0\,,
\label{G2F}
\ee
where $g_2(X)$ is a function of $X$. Then,
Eqs.~(\ref{qV}) and (\ref{cV}) reduce, respectively, to
\ba
q_V 
&=& 1+4\tilde{f}_6H^2 \phi^2\,,\label{qVest}\\
c_V^2
&=& \frac{1}{q_V} 
\left[ 1+\frac{(q_T-2B_4)^2}{2\phi^2 q_T} 
-2\tilde{f}_5 H\phi^3+\frac{\dot{\phi}}{H\phi} 
(q_V-1)\right]. 
\label{cvex}
\nonumber \\
\ea
For positive $\tilde{f}_6$, the vector ghost is absent. 
If $\tilde{f}_6$ is negative and the function $|\tilde{f}_6|H^2 \phi^2$ 
grows in time, there is a possibility for the appearance of 
ghosts. To avoid this, we require the condition
\be
|\tilde{f}_6| H^2 \phi^2 \ll 1
\label{f6con}
\ee
in the early cosmological epoch. 
Moreover, the condition $\tilde{f}_6H^2 \phi^2>-1/4$ 
needs to be satisfied on the de Sitter solution. 

For the theories with $\tilde{f}_5=0$, the vector propagation 
speed squared (\ref{cvex}) on the de Sitter solution 
($\dot{\phi}=0$) reduces to  
\be
(c_V^2)_{\rm dS}=\frac{1}{q_V} 
\left[ 1+\frac{(q_T-2B_4)^2}{2\phi^2 q_T} \right]\,,
\label{cvds}
\ee
which is positive under the no-ghost conditions 
$q_T>0$ and $q_V>0$. 
Provided that $q_V$ is close to 1 in the early cosmological 
epoch, the last term in the square brackets of Eq.~(\ref{cvex}) 
is also suppressed 
relative to the first term. 
Hence, the Laplacian instability 
can be avoided for the theories with 
$\tilde{f}_5=0$ and $\tilde{f}_6 \neq 0$.

In the presence of the coupling $\tilde{f}_5$, the term 
$-2\tilde{f}_5H \phi^3$ in Eq.~(\ref{cvex}) modifies the 
value of $c_V^2$. If $\tilde{f}_5$ is positive and 
the function $\tilde{f}_5H \phi^3$ grows in time, 
$c_V^2$ can be negative. To avoid this Laplacian 
instability, we require the condition 
\be
\tilde{f}_5H \phi^3 \ll 1
\label{f5con}
\ee
in the early cosmological epoch.
On the de Sitter solution, the term $\tilde{f}_5H \phi^3$ 
should not be large either to satisfy 
the stability condition $c_V^2>0$.

\begin{figure}
\begin{center}
\includegraphics[height=3.2in,width=3.3in]{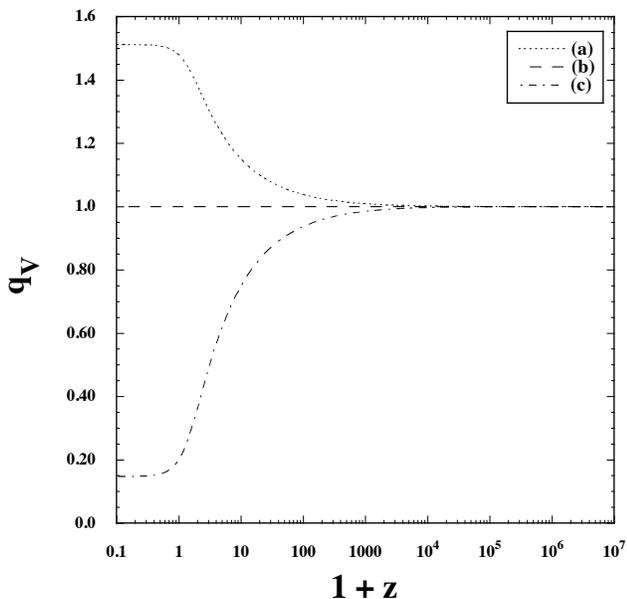}
\end{center}
\caption{\label{fig1}
Evolution of $q_V$ vs $1+z$ ($z$ is the redshift) 
in the presence of the 
coupling $\tilde{f}_6=c_6X^5$ for the three cases: 
(a) $\hat{c}_6 =3$, 
(b) $\hat{c}_6=0$, and (c) $\hat{c}_6=-5$. 
The functions $A_{2,3,4,5}$, which determine the background dynamics, 
are chosen as Eq.~(\ref{extendedco}) with 
$p_2=1$, $p=5$, $\beta_4=0.01$, $\beta_5=0.01$, 
$\lambda=1$, and $G_6=g_5=\tilde{f}_5=0$. 
The present epoch is identified by the 
condition $\Omega_{\rm DE}(z=0)=0.68$.
}
\end{figure}

\begin{figure}
\begin{center}
\includegraphics[height=3.2in,width=3.3in]{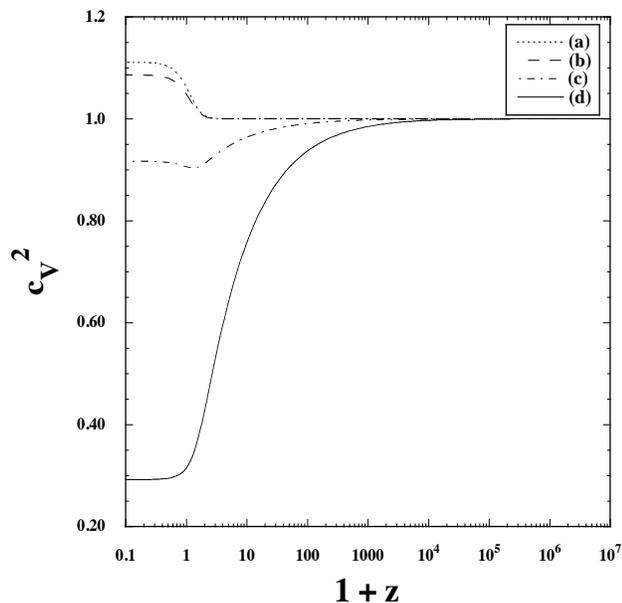}
\end{center}
\caption{\label{fig2}
Evolution of $c_V^2$ vs $1+z$ in the presence of 
the couplings $\tilde{f}_5=c_5 X^2$ and 
$\tilde{f}_6=c_6X^5$ for
$p_2=1$, $p=5$, $\beta_4=0.05$, $\beta_5=0.03$, 
$\lambda=1$, and $G_6=g_5=0$. 
The case (a) corresponds to the covariant EVG model 
with $\hat{c}_5=\hat{c}_6=0$,
whereas the other cases are the covariantized EVG model with 
(b) $\hat{c}_5=0$, $\hat{c}_6=0$, 
(c) $\hat{c}_5=0$, $\hat{c}_6=1$, and 
(d) $\hat{c}_5=1$, $\hat{c}_6=1$.
}
\end{figure}

\subsection{Covariantized EVG model}

For concreteness, let us consider the covariantized EVG model 
with the functions (\ref{G2F}) and the power-law couplings
\be
\tilde{f}_5(X)=c_5X^{q_5}\,,\qquad
\tilde{f}_6(X)=c_6X^{q_6}\,,
\label{f6}
\ee
where $c_5,c_6,q_5,$ and $q_6$ are constants. 
Since $\phi^p \propto H^{-1}$ for the background, 
we introduce the dimensionless constant 
\be
\lambda =\left( \frac{\phi}{M_{\rm pl}} \right)^p 
\frac{H}{m}\,,
\ee
where 
$m~(>0)$ is a mass scale related to the function $g_2(X)$ 
in Eq.~(\ref{G2F}) as $g_2(X)=b_2X^{p_2}$ 
with $b_2=-m^2M_{\rm pl}^{2(1-p_2)}$.
The negative value of $b_2$ is chosen to avoid 
the appearance of tensor ghosts in the limit 
that $G_5 \to 0$ \cite{DeFelice16}.
Then, the quantity $q_V$ reads 
\be
q_V=1+2^{2-q_6} \hat{c}_6 \lambda^2 
u^{2(1+q_6-p)}\,,
\label{qVexa}
\ee
where $\hat{c}_6=c_6 m^2 M_{\rm pl}^{2(1+q_6)}$, and 
\be
u \equiv \frac{\phi}{M_{\rm pl}}=
\left[ -2^{p_2} 
\frac{3\lambda^2 (p+p_2)\Omega_{\rm DE}}
{\gamma} \right]^{\frac{1}{2(p+p_2)}}\,.
\ee
For $q_6>p-1$, the function 
$u^{2(1+q_6-p)}$ in Eq.~(\ref{qVexa}) increases with 
the growth of $\phi$.

In Fig.~\ref{fig1}, we plot the evolution of $q_V$ 
for $q_6=5$, $p_2=1$, $p=5$, and $\lambda=1$ with three 
different values of $\hat{c}_6$. 
For $\hat{c}_6>0$, $q_V$ starts to grow from 
the value close to 1, and then it approaches a constant 
larger than 1 on the de Sitter attractor. 
For negative $\hat{c}_6$, $q_V$ decreases toward 
the value smaller than 1. 
Numerically, we find that the condition 
$\hat{c}_6 \gtrsim  -6$ is required for
avoiding the vector ghost.

For the theories with $\tilde{f}_5=0=\tilde{f}_6$, 
Eq.~(\ref{cvex}) reduces to $c_V^2=1+(q_T-2B_4)^2/(2\phi^2q_T)$, 
which is larger than 1 under the no-ghost 
condition $q_T>0$ of tensor perturbations. 
In case (a) of Fig.~\ref{fig2}, we plot the evolution 
of $c_V^2$ for the covariant EVG model with $B_4$ given by 
Eq.~(\ref{extendedco2}).
The vector propagation speed squared starts to evolve 
from the value close to 1, and then it finally approaches 
a superluminal value on the de Sitter attractor.
For the covariantized EVG model, the quantity $B_4$
is different, so $c_V^2$ exhibits some difference 
at late times compared to the 
covariant EVG model [see case (b) of Fig.~\ref{fig2}].

The case (c) of Fig.~\ref{fig2} corresponds to the covariantized 
EVG model for $\tilde{f}_5=0$ and 
$\tilde{f}_6=c_6X^5$ with $\hat{c}_6=1$. 
For $\hat{c}_6>0$ and $q_6>p-1$, the quantity $q_V$ 
grows toward a constant larger than 1 
[see case (a) of Fig.~\ref{fig1}].
This leads to the suppression of $c_V^2$ in Eq.~(\ref{cvds}) 
on the de Sitter solution. In case (c) of Fig.~\ref{fig2}, $c_V^2$ is 
in fact smaller than 1 at low redshifts. 
As we already mentioned, the theories with $\tilde{f}_5=0$ 
generally lead to the values of $(c_V^2)_{\rm dS}$ 
larger than zero.

In the covariantized EVG model, we have 
\be
\tilde{f}_5H \phi^3=2^{-q_5}\hat{c}_5\lambda 
u^{2q_5-p+3}\,,
\ee
where $\hat{c}_5=mM_{\rm pl}^{3+2q_5}c_5$. 
In case (d) of Fig.~\ref{fig2}, we show the evolution 
of $c_V^2$ for $q_5=2$, $\hat{c}_5=1$, 
$q_6=5$, $\hat{c}_6=1$, $p=5$, and $\lambda=1$, in
which case the quantity $\tilde{f}_5H \phi^3$ grows 
in proportion to $\phi^2$. 
Since $c_V^2 \simeq 0.29$ on the de Sitter solution, 
the Laplacian instability is absent.
In the covariantized EVG model 
studied in the numerical simulation 
of Fig.~\ref{fig2} ($\hat{c}_6=1$),
the condition $c_V^2>0$ is satisfied for 
$\hat{c}_5 \lesssim 1.5$. 

In summary, provided that the conditions (\ref{f6con}) 
and (\ref{f5con}) are satisfied for $\tilde{f}_6<0$ 
and $\tilde{f}_5>0$, the ghosts and Laplacian instabilities of 
vector perturbations do not generally arise from the BGP interactions
${\cal L}^{\rm N}$.

\section{Scalar perturbations}
\label{scasec}

\subsection{Stability conditions}
\label{scasec1}

Let us proceed to the discussion of no-ghost and stability 
conditions of scalar perturbations. The temporal and spatial 
components of the vector field contain the scalar perturbations 
$\delta \phi$ and $\chi_V$, respectively, as
\be
A^{0}=\phi(t)+\delta\phi\,,\qquad
A^{i}=\frac{1}{a^2(t)} \delta^{ij} \partial_{j}\chi_{V}\,.
\ee
We also consider the perturbed line element with scalar 
metric perturbations $\alpha$ and $\chi$ 
in the flat gauge, as
\be
ds_S^{2}=-(1+2\alpha)\,dt^{2}
+2\partial_i \chi dt\,dx^{i}
+a^{2}(t) \delta_{ij} dx^i dx^j\,. 
\label{smet}
\ee

If we consider two scalar fields $\sigma_r$ and $\sigma_m$ 
with kinetic terms $Z_r =-g^{\mu \nu} \partial_{\mu} \sigma_r \partial_{\nu} \sigma_r/2$ 
and $Z_m =-g^{\mu \nu} \partial_{\mu} \sigma_m \partial_{\nu} \sigma_m/2$ 
for the matter sector of scalar perturbations, then the k-essence action 
\be
S_M=\int d^4 x \sqrt{-g} \left[ P_r(Z_r)+P_m(Z_m) 
\right]\,
\ee
can describe the perfect fluids of radiation and 
nonrelativistic matter (labeled by $r$ and $m$, 
respectively) \cite{kes,Kase14}. 
At the background level, the fluid densities are 
$\rho_i=2Z_{i}P_{i,{Z}_i}-P_i$, where $i=r,m$. 
The density perturbation $\delta \rho_i$, 
the pressure perturbation $\delta P_i$, and the velocity 
potential $v_i$ are given, 
respectively, by \cite{HKT}
\ba
\delta \rho_i 
&=&\left( P_{i,Z_i}+2Z_iP_{i,Z_i Z_i} \right)\delta Z_i\,,
\label{corre1} \\
\delta P_i
&=& P_{i,Z_i} \delta Z_i\,,\\
v_i
&=& \frac{\delta \sigma_i}{\dot{\sigma}_i}\,,
\label{corre3}
\ea
where $\delta Z_i=\dot{\sigma}_i \delta \dot{\sigma}_i
-\dot{\sigma}_i^2\alpha$.

Expanding the action (\ref{action}) up to quadratic order 
in scalar perturbations, the second-order action reads \cite{HKT}
\ba
S_{S}^{(2)}  &=&  \int dt d^3 x\,a^{3}\,\Biggl\{
\left(w_{1}\alpha+\frac{w_{2} \delta\phi}{\phi} \right)\frac{\partial^{2}\chi}{a^2}-w_{3} 
\frac{(\partial \alpha)^{2}}{a^{2}} \nonumber\\
&&+w_{4}\alpha^{2}
-\frac{w_3}{4}\,\frac{(\partial\delta\phi)^2}{a^2 \phi^2}
+w_{5} \frac{(\delta\phi)^{2}}{\phi^2}
-\frac{w_{3}}{4\phi^{2}}\,\frac{(\partial\dot{\psi})^{2}}{a^{2}}
\nonumber \\
&&+\frac{w_{7}}{2}\,
\frac{(\partial\psi)^{2}}{a^{2}}-(3Hw_{1}-2w_{4})
\alpha \frac{\delta\phi}{\phi}  \nonumber \\
&&+\alpha\left[w_{3}\,\frac{\partial^{2}(\delta\phi)}
 {a^{2}\phi}+w_{3}\,\frac{\partial^{2}\dot{\psi}}{a^{2}\phi}
 -w_{6}\,\frac{\partial^{2}\psi}{a^{2}}\right]  \nonumber\\
&&-\left( w_8\psi-w_{3}\dot{\psi} \right)
 \frac{\partial^{2}(\delta\phi)}{2a^{2}\phi^{2}} \Biggr\} 
+(S_M)_S^{(2)}\,, 
\label{saction}
\ea
where $\psi \equiv \chi_V+\phi(t)\chi$, and 
\ba
w_{1} & = & -A_{3,X}\phi^2+4H(A_4-A_{4,X}\phi^2) \nonumber\\
&&+6H^2 (2A_5-A_{5,X}\phi^2)\,,
\label{w1}\\
w_{2} & = & w_1+2Hq_T\,,
\label{w2}\\
w_{3} & = & -2{\phi}^{2}q_V\,,
\label{w3} \\
w_{4} & = & 3H(w_2-Hq_T)+w_5\,,
\label{w4} \\
w_{5} & = & \frac12 \phi^4 \Big( A_{2,XX}
+3HA_{3,XX}  \nonumber\\
&&+6H^2A_{4,XX}+6H^3 A_{5,XX} \Big)\,,
\label{w5} \\
w_{6} & = & -\frac{1}{2\phi} 
\left[ 4H(q_T-2B_4)-w_8 \right]\,,
\label{w6} \\
w_{7} & = & \frac{2(q_T-2B_4)}{\phi^2} \dot{H}
+\frac{w_8}{2\phi^3}\dot{\phi}\,,
\label{w7}\\
w_{8} & = & 2w_2+4H\phi^2 \left( 2B_{4,X}
-H B_{5,X} \right)\,.
\label{w8}
\ea
The last term in Eq.~(\ref{saction}) corresponds to 
the second-order matter action 
$(S_M)_S^{(2)}=\int dtd^3x\,L_M$, with  
the Lagrangian 
\ba
L_M &=&
\sum_{i=r,m} a^3 \biggl[ \frac12 \left( P_{i,Z_i}
+\dot{\sigma}_i^2P_{i,{Z_iZ_i}} \right) 
\left( \dot{\delta \sigma_i}^2-2\dot{\sigma}_i \alpha 
\dot{\delta \sigma_i} \right) \nonumber \\
& &~\qquad \quad -\frac{1}{2a^2}P_{i,{Z_i}} \left\{ 
(\partial \delta \sigma_i)^2+2\dot{\sigma}_i
\partial \chi \partial \delta \sigma_i \right\} \nonumber \\
& &~\qquad \quad+\frac12 \dot{\sigma}_i^2 
\left( P_{i,Z_i}+\dot{\sigma}_i^2 P_{i,Z_iZ_i} 
\right) \alpha^2 \biggr]\,.
\ea
Varying the action (\ref{saction}) with respect to 
$\alpha, \chi, \delta \phi,$ and $\partial \psi,$ respectively, 
we obtain the perturbation equations of motion 
in Fourier space as
\ba
\hspace{-0.8cm}
&&
\sum_{i=r,m}\delta \rho_i-2w_4 \alpha+\left( 3Hw_1-2w_4 \right)
\frac{\delta \phi}{\phi}
\notag\\
\hspace{-0.8cm}
&&+\frac{k^2}{a^2} \left( {\cal Y}
+w_1 \chi-w_6 \psi \right)=0\,,
\label{per1} \\
\hspace{-0.8cm}
&&
\sum_{i=r,m} \left( \rho_i+P_i \right) v_i+
w_1 \alpha+\frac{w_2}{\phi} \delta \phi=0\,,
\label{per2}\\
\hspace{-0.8cm}
&&
\left( 3Hw_1-2w_4 \right)\alpha-2w_5 \frac{\delta \phi}{\phi}
\notag\\
\hspace{-0.8cm}
&&
+\frac{k^2}{a^2} \left( \frac12 {\cal Y}
+w_2 \chi-\frac{w_8}{2\phi} \psi
\right)=0\,,\label{per3} \\
\hspace{-0.8cm}
&&
\dot{\cal Y}+\left( H -\frac{\dot{\phi}}{\phi} \right){\cal Y}
+2\phi \left( w_6 \alpha+w_7 \psi \right)
+\frac{w_8}{\phi} \delta \phi
=0\,,\label{per4}
\ea
where
\be
{\cal Y} \equiv
 \frac{w_3}{\phi}
\left( \dot{\psi}+\delta \phi+ 2\phi \alpha \right)\,.
\label{Ydef}
\ee
The matter perturbation equations of motion, which 
follow from the continuity equations 
${\delta T^{\mu}_0}_{;\mu}=0$ and 
${\delta T^{\mu}_i}_{;\mu}=0$ for perturbations 
of the energy-momentum tensor 
$T^{\mu}_{\nu}=P_{i,{Z_i}}\partial ^{\mu}Z_i 
\partial _{\nu}Z_i+\delta^{\mu}_{\nu}P_i$, 
are given by 
\ba
\hspace{-0.8cm}
& &
\dot{\delta \rho}_i+3H \left( 1+c_i^2 \right) \delta \rho_i
+\frac{k^2}{a^2} \left( \rho_i+P_i \right)(\chi+v_i)=0\,, 
\label{per5} \\
\hspace{-0.8cm}
& &
\dot{v}_i-3Hc_i^2v_i-c_i^2 \frac{\delta \rho_i}{\rho_i+P_i} 
-\alpha=0\,,
\label{per6}
\ea
where $i=r,m$, and 
\be
c_i^2 \equiv \frac{P_{i,Z_i}}{\rho_{i,Z_i}}
=\frac{P_{i,Z_i}}{P_{i,Z_i}+2Z_iP_{i,Z_iZ_i}}\,.
\ee

By using Eqs.~(\ref{per1})--(\ref{per3}) with 
Eqs.~(\ref{corre1})--(\ref{corre3}), one can express 
$\alpha, \chi, \delta \phi$ in terms of 
$\psi, \delta \sigma_r, \delta \sigma_m$ and 
their derivatives. Then, the second-order action 
(\ref{saction}) can be expressed in the form 
$S_{S}^{(2)}=\int dt d^3x\,L$, 
with the Lagrangian
\be
L=a^{3}\Big( \dot{\vec{\mathcal{X}}}^{t}{\bm K}
\dot{\vec{\mathcal{X}}}
+\frac{k^2}{a^2}\vec{\mathcal{X}}^{t}{\bm G}
\vec{\mathcal{X}} 
 -\vec{\mathcal{X}}^{t}{\bm M}
\vec{\mathcal{X}}
-\vec{\mathcal{X}}^{t}{\bm B}
\dot{\vec{\mathcal{X}}}
\Big) \,,
\label{SSfinal}
\ee
where ${\bm K}$, ${\bm G}$, ${\bm M}$, ${\bm B}$ are $3 \times 3$ matrices 
and $\vec{\mathcal{X}}^{t}=\left( \psi, \delta \sigma_r, \delta \sigma_m\right)$. 
In the small-scale limit, the nonvanishing components of 
the matrices ${\bm K}$ and ${\bm G}$ are given by
\ba
K_{11}&=&Q_S+\xi_{r1}^2 K_{22}+\xi_{m1}^2 K_{33}\,, \nonumber\\
K_{22}&=&\frac12 \left( P_{r,Z_r}+\dot{\sigma}_r^2P_{r,Z_rZ_r} \right)\,,\nonumber\\
K_{33}&=&\frac12 \left( P_{m,Z_m}+\dot{\sigma}_m^2P_{m,Z_mZ_m} \right)\,,\nonumber\\
K_{12}&=&K_{21}=\xi_{r1} K_{22}\,,\notag\\
K_{13}&=&K_{31}=\xi_{m1} K_{33}\,,
\ea
and 
\ba
G_{11}&=&{\cal G}+\dot{\mu}+H\mu\,,\nonumber\\
G_{22}&=&\frac12 P_{r,Z_r}\,,\nonumber\\
G_{33}&=&\frac12 P_{m,Z_m}\,,\nonumber\\
G_{12}&=&G_{21}=\xi_{r2} G_{22}\,,\notag\\
G_{13}&=&G_{31}=\xi_{m2} G_{33}\,,
\ea
with 
\ba
&&Q_S=\frac{H^2q_T(3w_1^2+4q_Tw_4)}
{(w_1-2w_2)^2\phi^2}\,,\nonumber\\
&&\xi_{r1}=-\frac{w_2\dot{\sigma}_r}{(w_1-2w_2)\phi}\,, \quad 
\xi_{m1}=-\frac{w_2\dot{\sigma}_m}{(w_1-2w_2)\phi}\,,
\nonumber\\
&&\xi_{r2}=-\frac{(w_8-w_6\phi)\dot{\sigma}_r}{(w_1-2w_2)\phi}\,, \quad 
\xi_{m2}=-\frac{(w_8-w_6\phi)\dot{\sigma}_m}{(w_1-2w_2)\phi}\,,
\nonumber \\
&&{\cal G}=\frac{w_1w_8(4w_2w_6\phi-w_1w_8)-4w_2^2w_6^2 \phi^2}
{4w_3(w_1-2w_2)^2\phi^2}-\frac{w_7}{2},
\nonumber \\
&&\mu=\frac{2w_2w_6\phi-w_1w_8}{4(w_1-2w_2)\phi^2}\,.
\ea
Under the no-ghost conditions $K_{22}>0$ and $K_{33}>0$ of the matter fields, 
the positivity of ${\bm K}$ is ensured for $Q_S>0$. 

The scalar propagation speeds $c_S$ are the solutions 
to the dispersion relation given by 
${\rm det}\left( c_S^2 {\bm K}-{\bm G} \right)=0$, i.e., 
\ba
&&\left(c_S^2 K_{11}-G_{11} \right)
\left(c_S^2 K_{22}-G_{22} \right)
\left(c_S^2 K_{33}-G_{33} \right)\notag\\
&&-\left(c_S^2 K_{12}-G_{12} \right)^2 \left(c_S^2 K_{33}-G_{33} \right)\notag\\
&&-\left(c_S^2 K_{13}-G_{13} \right)^2 \left(c_S^2 K_{22}-G_{22} \right)=0\,.
\label{csso}
\ea
It is useful to notice the following relation, 
\be
w_8-(w_6\phi+w_2)=-4H\phi^4 \left( f_4+3H\phi f_5 \right)\,,
\ee
where we used Eqs.~(\ref{ABrelation0}) and (\ref{ABrelation}).
Since $f_4=0=f_5$ in GP theories, we have that 
$w_8=w_6\phi+w_2$. 
Provided that $f_4=0=f_5$, the same relation holds even 
in BGP theories with 
nonvanishing functions $\tilde{f}_5, \tilde{f}_6$.
In such cases, we have $\xi_{r1}=\xi_{r2}$, $\xi_{m1}=\xi_{m2}$, 
so that $K_{12}/K_{22}=G_{12}/G_{22}$ 
and $K_{13}/K_{33}=G_{13}/G_{33}$.
Then, Eq.~(\ref{csso}) gives 
the three decoupled solutions
\ba
c_r^2 &=&\frac{G_{22}}{K_{22}}\,,
\label{cr} \\
c_m^2 &=&\frac{G_{33}}{K_{33}}\,,
\label{cm} \\
c_{\rm P}^2 &=& \frac{1}{Q_S} \left(
G_{11}-\xi_{r1}^2 G_{22}-\xi_{m1}^2G_{33}\right)\,,
\label{cp}
\ea
where $c_{\rm P}$ corresponds to the scalar propagation 
speed arising from the longitudinal mode of the vector field.

In BGP theories with nonvanishing 
functions $f_4$ and $f_5$, the three propagation speeds 
are mixed with each other. To quantify the deviation 
from GP theories in the scalar sector, 
we define the following quantities: 
\ba
\alpha_{\rm P} &\equiv& \frac{\xi_{r2}}{\xi_{r1}}-1
=\frac{w_8-(w_6\phi+w_2)}{w_2}\,,
\label{alphaP}\\
\beta_{{\rm P}r} &\equiv& \frac{2\xi_{r1}^2 G_{22} \alpha_{\rm P}}{Q_S} \nonumber\\
&=&\frac{w_2(w_8-w_6\phi-w_2)(\rho_r+P_r)}
{(3w_1^2+4q_Tw_4)q_TH^2}\,,\\
\beta_{{\rm P}m} &\equiv& \frac{2 \xi_{m1}^2 G_{33}\alpha_{\rm P}}{Q_S} \nonumber\\
&=&\frac{w_2(w_8-w_6\phi-w_2)(\rho_m+P_m)}
{(3w_1^2+4q_Tw_4)q_TH^2}\,.
\label{bepr}
\ea
In the limit that $c_m^2 \to 0$, one of the solutions to 
Eq.~(\ref{csso}) is given by $c_S^2=0$, whereas 
the other two solutions are 
\be
c_S^2=
\frac12 \left[ c_r^2+c_{\rm P}^2-\beta_{\rm P}
\pm \sqrt{(c_r^2-c_{\rm P}^2+\beta_{\rm P})^2+
2c_r^2 \alpha_{\rm P} \beta_{{\rm P}r}} \right],
\label{cS2}
\ee
where $c_{\rm P}^2$ is of the same form as Eq.~(\ref{cp}), i.e., 
\be
c_{\rm P}^2=\frac{1}{Q_S} \left[ {\cal G}+
\dot{\mu}+H\mu -\frac{w_2^2 (\rho_r+P_r+\rho_m+P_m)}{2(w_1-2w_2)^2\phi^2} \right],
\label{cP}
\ee
and 
\be
\beta_{\rm P}=\beta_{{\rm P}r}+\beta_{{\rm P}m}\,.
\ee

If the deviation from GP theories is small, 
then the contribution $2c_r^2 \alpha_{\rm P} \beta_{{\rm P}r}$ 
to $c_S^2$ should be subdominant to the term 
$(c_r^2-c_{\rm P}^2+\beta_{\rm P})^2$ in Eq.~(\ref{cS2}).
In this case, one of the solutions to Eq.~(\ref{cS2}) 
reduces to $c_S^2 \simeq c_r^2$, while 
another solution reads
\be
c_{\rm S}^2 \simeq c_{\rm P}^2-\beta_{\rm P}\,.
\label{cscpbe}
\ee
Thus, the deviation from GP theories 
($\beta_{\rm P} \neq 0$) in the scalar sector leads to the value of 
$c_{\rm S}^2$ different from $c_{\rm P}^2$.
The Laplacian instability can be avoided for $c_{\rm S}^2>0$. 
The sound speeds derived above are the generalizations of 
the single-fluid case discussed in Ref.~\cite{HKT}.

\subsection{Covariantized EVG model}

We compute the quantities $Q_S$ and $c_{\rm S}^2$ for the 
covariantized EVG model to discuss theoretically 
viable parameter spaces.
Under the no-ghost condition $q_T>0$ of tensor 
perturbations, we require that the quantity 
$q_S \equiv 3w_1^2+4q_Tw_4$ in $Q_S$ is positive. 
This amounts to the condition 
\ba
\hspace{-0.9cm}
& &
q_S=
-2^{2-p_2}b_2 p_2
\left( p+ p_2\Omega_{\rm DE} \right)
M_{\rm pl}^{2(1+p_2)}u^{2p_2} \nonumber \\
\hspace{-0.9cm}
& &\qquad \times
\left[ 1-6(2p+2p_2-1)\beta_4+2(3p+2p_2)\beta_5
\right]>0.
\label{qsex}
\ea
In the limit that $|\beta_4| \ll 1$ and $|\beta_5| \ll 1$, 
the condition (\ref{qsex}) is satisfied for $b_2<0$ 
with positive values of $p_2, p$,  and $u=\phi/M_{\rm pl}$.
Even for $q_S>0$, there are cases in which the term 
$w_1-2w_2$ in the denominator of $Q_S$ 
crosses zero \cite{DeFelice16}. 
The quantity $w_1-2w_2$ can be expressed as 
\be
w_1-2w_2=-2HM_{\rm pl}^2 
\left( 1-\Omega_{\rm DE}w_c \right)\,,
\label{w12}
\ee
where $w_c \equiv 1+p+p_2-(p+1)(p+p_2)
(2\beta_5 p_2-1)/\gamma$. Provided that the dark energy 
density parameter is in the range $0<\Omega_{\rm DE}<1$, 
the rhs of Eq.~(\ref{w12}) remains negative for $w_c<1$, i.e.,
\be
\frac{(p+1) (2\beta_5 p_2-1)}{\gamma}>1\,.
\label{divcon}
\ee
For the theories with $\beta_5=0$ and $p>-1$, the condition 
$\gamma<0$ is necessary to satisfy Eq.~(\ref{divcon}).

In the covariantized EVG model, the quantity $\beta_{\rm P}$ 
arising from the deviation from GP theories yields
\be
\beta_{\rm P}=\frac{2\Omega_{\rm DE}
[3(1-\Omega_{\rm DE})+\Omega_r]}
{p+p_2\Omega_{\rm DE}}
\frac{{\cal A}_1}{{\cal A}_2}\,,
\ee
where 
\ba
{\cal A}_1 
&=& p_2 
[\beta_4 (1-2p-2p_2)+\beta_5(p+p_2)]\,,\\
{\cal A}_2 
&=& p+2p p_2
[\beta_5 (2-3\Omega_{\rm DE})-6\beta_4(1-\Omega_{\rm DE})] \nonumber \\
& &+p_2[1+2p_2 \beta_5(2-3\Omega_{\rm DE}) 
\nonumber \\
& &
+6\beta_4(1-\Omega_{\rm DE})(1-2p_2)]\,.
\ea
The quantity $\beta_{\rm P}$ vanishes in the limit that 
$\Omega_{\rm DE} \to 0$. 
Moreover, we also have $\beta_{\rm P} \to 0$ in the 
de Sitter limit ($\Omega_{\rm DE} \to 1$ and 
$\Omega_r \to 0$). Hence, the quantity $\beta_{\rm P}$ can 
deviate from zero only during the transition from the matter 
era to the de Sitter epoch.

During the radiation, deep matter, and de Sitter epochs, we 
can employ the approximation $c_{\rm S}^2 \simeq c_{\rm P}^2$ 
in Eq.~(\ref{cscpbe}), so the corresponding value of $c_{\rm S}^2$ 
in each cosmological epoch reads
\begin{widetext}
\ba
(c_{\rm S})_r^2
& &=\frac{1}{3p^2} \left[ 4p_2-2+
\frac{3p \{ 2(2p+2p_2-1)(2-3p-3p_2)\beta_4+
(p+p_2)[1-(4-6p-4p_2)\beta_5]\}}
{(p+p_2)[1-6(2p+2p_2-1)\beta_4+(6p+4p_2)\beta_5]} 
\right]\,,\label{csr} \\
(c_{\rm S})_m^2
& &=\frac{1}{6p^2} \left[ 6p_2-3+
\frac{p \{ 6(2p+2p_2-1)(3-5p-5p_2)\beta_4+
(p+p_2)[5-2(9-15p-10p_2)\beta_5]\}}
{(p+p_2)[1-6(2p+2p_2-1)\beta_4+(6p+4p_2)\beta_5]} 
\right]\,,\label{csm}\\
(c_{\rm S})_{\rm dS}^2
& &=\frac{2p_2[1-6(2p+2p_2-1)\beta_4+(6p+4p_2)\beta_5]
+\{ 1-p_2[1-6(2p+2p_2-1)\beta_4+2(1+3p+2p_2)\beta_5]\}(q_Vu^2)_{\rm dS}}
{3(p+p_2)(1-2p_2 \beta_5)(q_V u^2)_{\rm dS}}. 
\nonumber \\
\label{csds}
\ea
\end{widetext}
In the limit that $|\beta_4| \ll 1$ and $|\beta_5| \ll 1$, 
Eqs.~(\ref{csr})--(\ref{csds}) reduce to 
\ba
(c_{\rm S})_r^2
& &=\frac{3p+4p_2-2}{3p^2}\,,
\label{csrl}\\
(c_{\rm S})_m^2
& &=\frac{5p+6p_2-3}{6p^2}\,,\\
(c_{\rm S})_{\rm dS}^2
& &=\frac{2p_2+(1-p_2)(q_V u^2)_{\rm dS}}{3(p+p_2)(q_V u^2)_{\rm dS}}\,.
\label{csdsl}
\ea
The two stability conditions $(c_{\rm S})_r^2>0$ and 
$(c_{\rm S})_m^2>0$ are ensured for 
\ba
& &
3p+4p_2-2>0 \qquad ({\rm if}~p_2<1/2),\label{stacon1}\\
& &
5p+6p_2-3>0 \qquad ({\rm if}~p_2>1/2).
\ea
For positive integers $p$ and $p_2$, these conditions 
are trivially satisfied. If we demand the absence of 
Laplacian instabilities on the de Sitter attractor, 
we require that $(c_{\rm S})_{\rm dS}^2>0$. 
For $p+p_2>0$, this condition translates to 
\be
2p_2+(1-p_2)(q_V u^2)_{\rm dS}>0\,,\label{stacon3}
\ee
where we used the no-ghost condition of vector perturbations. 
In the limit that $q_V \to \infty$, the de Sitter stability 
is ensured for $p_2 \leq1$, whereas, in another limit 
$q_V \to 0$, the de Sitter solution is stable for any 
positive value of $p_2$. 

\begin{figure}
\begin{center}
\includegraphics[height=3.2in,width=3.2in]{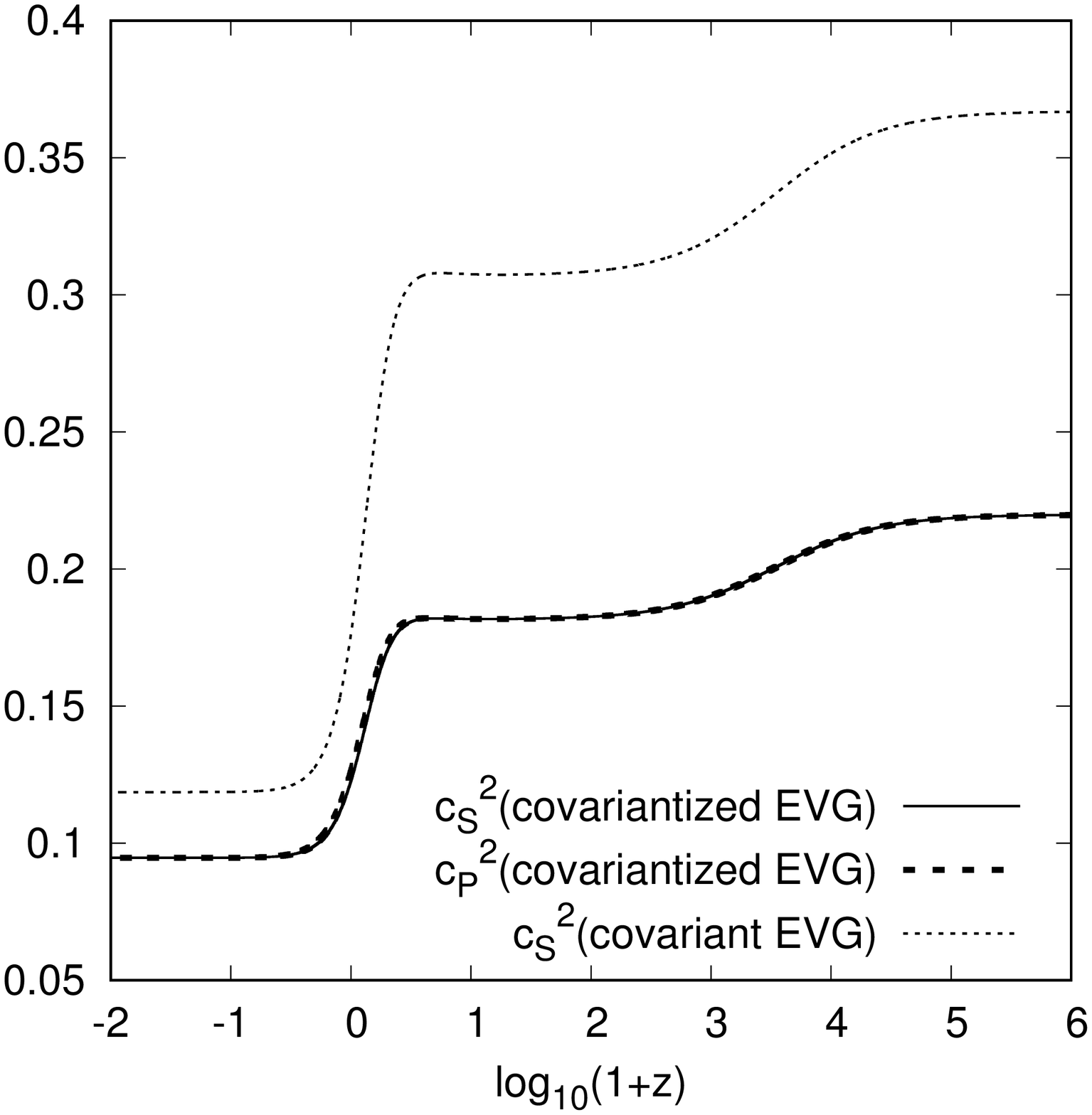}
\vspace{0.2cm}

\includegraphics[height=3.2in,width=3.2in]{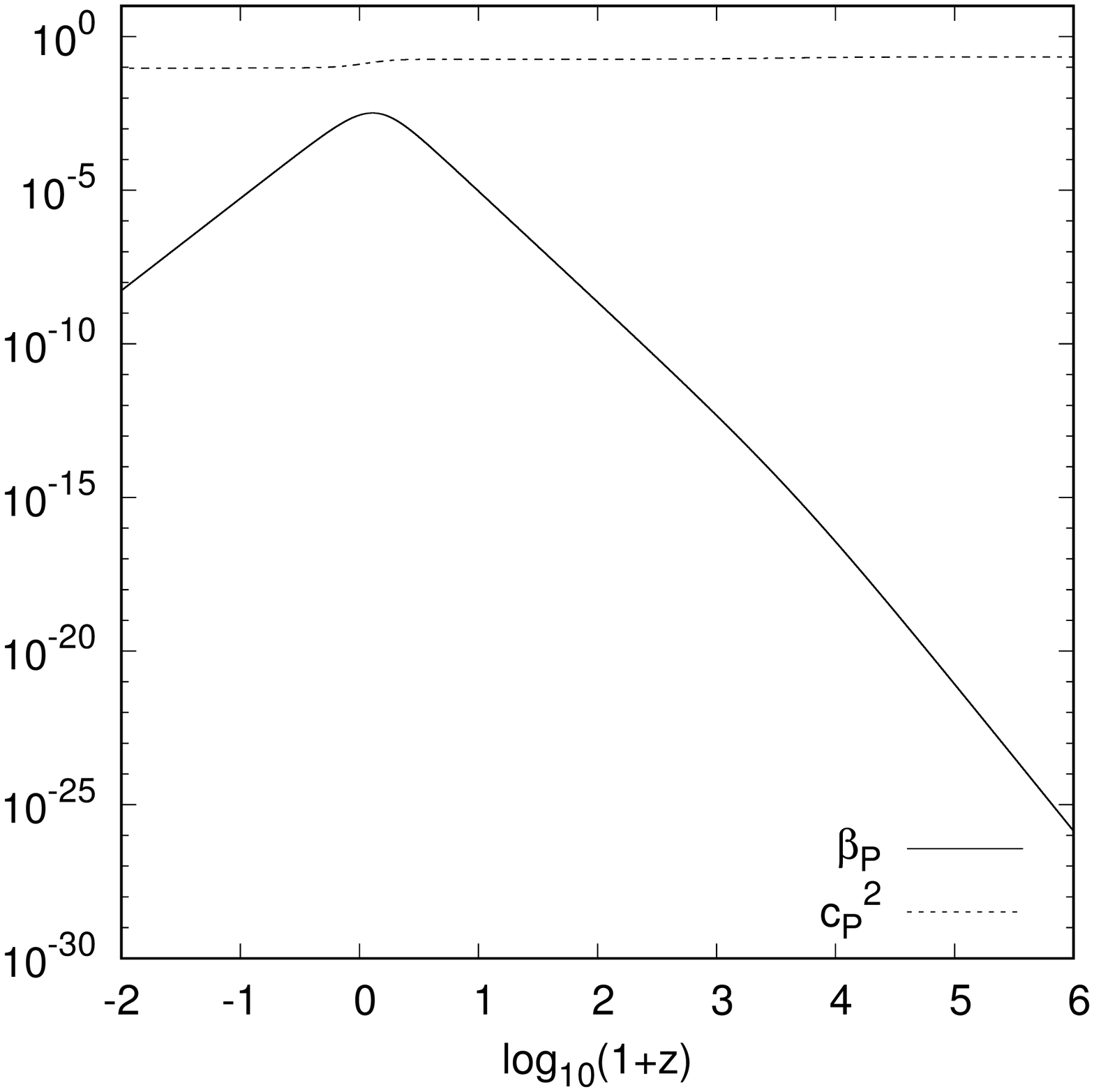}
\end{center}
\caption{\label{fig3}
(Top) Evolution of $c_{\rm S}^2$ versus $\log_{10}(1+z)$ 
for the covariantized EVG model (thin solid line) and 
the covariant EVG model (thin dotted line)
with the model parameters $p_2=1$, $p=5$, $\beta_4=0.01$, 
$\beta_5=0.03$, $\lambda=1$, and $g_5=G_6=\tilde{f}_5=\tilde{f}_6=0$. 
The variation of $c_{\rm P}^2$ in the covariantized EVG model is also plotted 
as a bold dotted line. (Bottom) Evolution of $\beta_{\rm P}$ and $c_{\rm P}^2$ 
versus $\log_{10}(1+z)$ for the covariantized EVG model 
with the same model parameters used above.
Since $c_{\rm P}^2 \gg \beta_{\rm P}$, $c_{\rm S}^2$ 
in the covariantized EVG model is very close to $c_{\rm P}^2$ (see the top panel).  
The difference of $c_{\rm S}^2$ between covariantized 
and covariant EVG models mostly comes from the 
difference of $c_{\rm P}^2$ between the two models.}
\end{figure}

In the covariant EVG model, the quantity $\beta_{\rm P}$ 
vanishes, so $c_{\rm S}^2$ is exactly equivalent to 
$c_{\rm P}^2$. Since $c_{\rm P}^2$ contains the functions 
that depend on $B_4,B_5$ (like $w_6,w_7,w_8$), the value of 
$c_{\rm P}^2$ in the covariant EVG model is different from 
that in the covariantized EVG model. 
In the Appendix we show the values of $c_{\rm S}^2$ 
in the covariant EVG model during the radiation and early 
matter eras ($\Omega_{\rm DE} \to 0$) as well as
during the de Sitter epoch ($\Omega_{\rm DE} \to 1, \Omega_r \to 0$); 
see Eqs.\,(\ref{csr_covariant})--(\ref{csds_covariant}).
They are indeed different from Eqs.\,(\ref{csr})--(\ref{csds}), 
but in the limit that $|\beta_4| \ll 1$ and $|\beta_5| \ll 1$, 
they reduce to the values (\ref{csrl})--(\ref{csdsl}), 
respectively. This means that the difference of $c_{\rm S}^2$ 
between the two models mostly arises from the different choices 
of the functions $B_4$ and $B_5$ in $c_{\rm P}^2$.
Apart from the transient period from the matter era to 
the de Sitter epoch, the quantity $\beta_{\rm P}$ is close 
to zero in the covariantized EVG model, so the contribution of 
$\beta_{\rm P}$ to 
Eq.~(\ref{cscpbe}) should be small relative to $c_{\rm P}^2$.

To confirm the above analytic estimation, we numerically 
compute $c_{\rm S}^2$ for $|\beta_4|$ and $|\beta_5|$ 
smaller than the order of 1. In the top panel of 
Fig.~\ref{fig3}, the evolution of $c_{\rm S}^2$ for $p_2=1$, $p=5$, 
$\beta_4=0.01$, $\beta_5=0.03$, and $\lambda=1$ 
is plotted in both covariantized and covariant EVG models 
with vanishing functions $g_5,G_6,\tilde{f}_5,\tilde{f}_6$.
In this case, the stability conditions 
(\ref{stacon1})--(\ref{stacon3}) are automatically satisfied. 
The numerical values of $c_{\rm S}^2$ 
exhibit excellent agreement with Eqs.\,(\ref{csr})--(\ref{csds}) 
in the covariantized EVG model and 
Eqs.\,(\ref{csr_covariant})--(\ref{csds_covariant})
in the covariant EVG model. 

In the top panel of Fig.~\ref{fig3}, we also show the 
evolution of $c_{\rm P}^2$ in the covariantized EVG model 
as a bold dotted line. The value of $c_{\rm S}^2$ in this 
model is almost identical to $c_{\rm P}^2$ apart from 
the tiny deviation around today. As we see in the bottom 
panel of Fig.~\ref{fig3}, the quantity $\beta_{\rm P}$ has 
a peak around $z=0$ with the asymptotic behavior 
$\beta_{\rm P} \to 0$ in the past and the future.
Since the condition $c_{\rm P}^2 \gg \beta_{\rm P}$ 
always holds during the cosmic expansion history, 
$c_{\rm S}^2$ is practically identical to $c_{\rm P}^2$ 
in the covariantized EVG model. As we see in the 
top panel of Fig.~\ref{fig3}, the value of $c_{\rm S}^2$
in the covariantized EVG model differs from that in 
the covariant EVG model. This is mostly attributed to the 
difference of $c_{\rm P}^2$ between the two models. 

\section{Matter density perturbations and gravitational potentials}
\label{effsec}

To confront BGP theories with the observations of large-scale structures 
and weak lensing, we need to study the evolution of matter density perturbations 
and gravitational potentials. 
For this purpose, we define the gauge-invariant 
density contrast $\delta$ of nonrelativistic matter 
(satisfying $w=0$ and $c_m^2=0$) as
\be
\delta \equiv \frac{\delta \rho_m}{\rho_m}
+3Hv\,.
\ee
We also introduce the gauge-invariant gravitational 
potentials \cite{Bardeen}
\be
\Psi \equiv \alpha+\dot{\chi}\,,\qquad
\Phi \equiv H \chi\,
\ee
and the gravitational slip parameter 
\be
\eta \equiv -\frac{\Phi}{\Psi}\,.
\label{slip}
\ee
Taking the time derivative of Eq.~(\ref{per5}) and using Eq.~(\ref{per6}), 
the density contrast of nonrelativistic matter obeys 
\be
\ddot{\delta}+2H\dot{\delta}+\frac{k^2}{a^2}\Psi
=3 \ddot{\cal B}+6H \dot{\cal B}\,,
\label{deltaeq}
\ee
where ${\cal B} \equiv Hv$.
We also express the relation between $\Psi$ and $\delta$ 
as a form of the modified Poisson equation 
\be
\frac{k^2}{a^2}\Psi=-4\pi G_{\rm eff} \rho_m \delta\,.
\ee
The effective gravitational coupling $G_{\rm eff}$, 
which is a key quantity that determines the growth rate of 
matter perturbations according to Eq.~(\ref{deltaeq}), 
is known by solving the 
other perturbation equations of motion. 

Another important quantity associated with the deviation 
of light rays in weak lensing observations 
is given by \cite{weaklensing}
\be
\Phi_{\rm eff} \equiv \Phi-\Psi
=-(\eta+1)\Psi\,.
\label{Phieff}
\ee
In General Relativity, the gravitational slip parameter (\ref{slip}) is 
equivalent to 1 in the absence of the anisotropic stress, 
so that $\Phi_{\rm eff}=-2\Psi=2\Phi$. 
In BGP theories, the quantity $\eta$ generally 
varies in time at low redshifts, so it affects the evolution 
of $\Phi_{\rm eff}$.

\subsection{Quasistatic approximation for 
subhorizon perturbations} 

To test for BGP theories with the observations 
of large-scale structures and weak lensing, 
we are primarily interested in the evolution of 
nonrelativistic matter perturbations 
for the modes deep inside the Hubble radius. 
As long as the oscillating mode of a scalar degree of 
freedom is negligible relative to the matter-induced mode, 
it is known that the so-called quasistatic 
approximation \cite{Polarski} is 
sufficiently accurate for perturbations deep inside 
the sound horizon ($c_{\rm S}^2\,k^2/a^2 \gg H^2$) in 
Horndeski theories \cite{DKT} and generalized Proca 
theories \cite{DeFelice16}. 
Under this approximation scheme, the dominant 
contributions to the perturbation equations of motion are 
those containing the matter perturbation $\delta \rho_m$ and the term $k^2/a^2$. 
We assume that $c_{\rm S}^2$ is not very close 
to zero, so that the condition $c_{\rm S}^2\,k^2/a^2 \gg H^2$ holds 
for perturbations associated with observed large-scale structures.

We employ the quasistatic approximation explained above 
without taking into account the radiation. 
Then, Eqs.~(\ref{per1}) and (\ref{per3}) 
reduce, respectively, to 
\ba
& &
\delta \rho_m+\frac{k^2}{a^2} \left( 
{\cal Y}+w_1 \chi -w_6 \psi \right) 
\simeq 0\,,\label{qua1}\\
& &
{\cal Y} \simeq -2w_2 \chi+\frac{w_8}{\phi} 
\psi\,,\label{qua2}
\ea
so we obtain
\be
\delta \rho_m \simeq -\frac{k^2}{a^2} \left[ 
\frac{w_1-2w_2}{H}\Phi
+\frac{w_2}{\phi} \left(1+\alpha_{\rm P} \right)
\psi \right]\,,
\label{quasieq1}
\ee
where $\alpha_{\rm P}$ is defined by Eq.~(\ref{alphaP}).
Eliminating the velocity potential $v_m$ from 
Eqs.~(\ref{per2}) and (\ref{per5}), 
it follows that 
\be
\dot{\delta \rho}_m+3H\delta \rho_m
+\frac{k^2}{a^2} \left( \frac{\rho_m}{H}\Phi 
-w_1 \alpha -\frac{w_2}{\phi} \delta \phi 
\right)=0\,.
\label{qua3}
\ee
We take the time derivative of Eq.~(\ref{qua1}) 
and eliminate the terms $\dot{\delta \rho}_m$ and 
$\delta \rho_m$ from Eq.~(\ref{qua3}).
In doing so, we use the definition of ${\cal Y}$ 
with Eq.~(\ref{qua2}) to remove the perturbation 
$\delta \phi$. This leads to the following equation, 
\be
\phi^2 (w_1-2w_2)w_3 \Psi+\mu_1\Phi 
+\mu_2\psi+\aP\phi w_2w_3\dot{\psi}
\simeq 0\,,
\label{quasieq2}
\ee
where 
\ba
\mu_1 
&=& \frac{\phi^2}{H} [ w_3(\dot{w}_1-2\dot{w}_2+Hw_1-\rho_m) \nonumber \\
& &-2w_2 (w_2+Hw_3)]\,,\label{mu1}\\
\mu_2
&=& 
\phi^2w_2w_6+(1+\aP)[\phi(w_2^2+Hw_2w_3+\dot{w_2}w_3)\nonumber \\
& &
-\dot{\phi}w_2w_3]+\dot{\alpha}_{\rm P} \phi w_2w_3\,. 
\ea
Differentiating Eq.~(\ref{qua2}) with respect to $t$ 
and eliminating the terms $\dot{\cal Y}$ and ${\cal Y}$ 
from Eq.~(\ref{per4}), it follows that 
\be
2\phi^2 w_2 (1+\alpha_{\rm P}) \Psi
+\mu_3\Phi +\mu_4 \psi-\frac{2\alpha_{\rm P}\phi^2 w_2}{H} 
\dot{\Phi} \simeq 0\,,
\label{quasieq3}
\ee
where 
\ba
\mu_3
&=&
\frac{2\phi}{Hw_3}\Big[ \phi^2 w_2w_6+\phi(w_2^2+Hw_2w_3+\dot{w_2}w_3)
-\dot{\phi}w_2w_3\notag\\
&&
+\frac{\aP\phi w_2}{H}(Hw_2+\dot{H}w_3)\Big]\,,\\
\mu_4
&=& -\frac{1}{w_3}[\phi^3(2w_3w_7+w_6^2)
+\phi^2\{Hw_3w_6+w_3\dot{w_6}\notag\\
&&+2(1+\aP)w_2w_6\}
+\phi\{(1+\aP)(Hw_2w_3+\dot{w_2}w_3)\notag\\
&&+\dot{\alpha}_{\rm P}w_2w_3-\dot{\phi}w_3w_6+(1+\aP)^2w_2^2\}\notag\\
&&-2(1+\aP)\dot{\phi}w_2w_3]
\,.\label{mu4}
\ea

In GP theories, we have $\alpha_{\rm P}=0$, 
in which case the two terms containing the time derivatives 
$\dot{\psi}$ and $\dot{\Phi}$ vanish in 
Eqs.~(\ref{quasieq2}) and (\ref{quasieq3}). 
Then, the three equations (\ref{quasieq1}), (\ref{quasieq2}), 
and (\ref{quasieq3}) are closed, so they can be explicitly 
solved for $\Psi, \Phi$, and $\psi$ \cite{Geff}.
This property does not hold for BGP theories
with a nonvanishing value of $\alpha_{\rm P}$.
In this case, we need to deal with Eqs.~(\ref{quasieq2}) 
and (\ref{quasieq3}) as first-order differential equations. 
Let us introduce the dimensionless quantities
\be
\epsilon_{\psi}=\frac{\dot{\psi}}{H\psi}\,,\qquad
\epsilon_{\Phi}=\frac{\dot{\Phi}}{H\Phi}\,.
\ee
On using Eqs.~(\ref{quasieq1}), (\ref{quasieq2}), 
and (\ref{quasieq3}), we can express $\Psi, \Phi,$ and $\psi$ 
in the following forms:
\ba
\Psi &\simeq& -\frac{{\cal F}_1}{\phi \mu_5} 
\frac{a^2}{k^2} \delta \rho_m\,,\label{Psiqu}\\
\Phi &\simeq& \frac{{\cal F}_2}{\mu_5} 
\frac{a^2}{k^2} \delta \rho_m\,,\label{Phiqu}\\
\psi &\simeq& \frac{{\cal F}_3}{\mu_5} 
\frac{a^2}{k^2} \delta \rho_m\,,\label{psiqu}
\ea
where
\ba
\hspace{-.2cm}
\mu_5 
&=&
-(1+\aP) H w_2 [(w_1-2w_2)w_3\mu_3-2(1+\aP)w_2\mu_1]\notag\\
\hspace{-.2cm}
&&+\phi(w_1-2w_2)[(w_1-2w_2)w_3\mu_4-2(1+\aP)w_2\mu_2]\notag\\
\hspace{-.2cm}
&&+2\aP (1+\aP) \phi^2 H w_2^2w_3(w_1-2w_2)(\epsilon_{\Phi}-\epsilon_{\psi}),
\label{mu5}\\
\hspace{-.2cm}
{\cal F}_1
&=&
H[\mu_2\mu_3-\mu_1\mu_4+\aP\phi w_2 \{Hw_3\mu_3\epsilon_{\psi}\notag\\
\hspace{-.5cm}
&&-2\phi\epsilon_{\Phi} (\mu_2+\aP\phi H w_2w_3\epsilon_{\psi})\}]
\,,\\
\hspace{-.5cm}
{\cal F}_2
&=&
\phi H [2 (1+\aP) w_2 (\mu_2+\aP\phi H w_2w_3\epsilon_{\psi})\notag\\
\hspace{-.5cm}
&&-(w_1-2w_2)w_3\mu_4]
\,,\\
\hspace{-.5cm}
{\cal F}_3
&=&
\phi H [(w_1-2w_2)w_3 (\mu_3-2\aP\phi^2w_2\epsilon_{\Phi})\notag\\
\hspace{-.5cm}
&&-2 (1+\aP) w_2 \mu_1]
\,.
\label{F3}
\ea
For $\alpha_{\rm P} \neq 0$, Eqs.~(\ref{Psiqu})--(\ref{psiqu}) are not closed, 
so we need to solve the other perturbation equations of motion to find the evolution of 
$\epsilon_{\psi}$ and 
$\epsilon_{\Phi}$ for a given model.

In BGP theories with nonvanishing $\tilde{f}_5, \tilde{f}_6$ 
but with vanishing $f_4, f_5$, the quantity $\alpha_{\rm P}$ 
vanishes, so Eqs.~(\ref{Psiqu})--(\ref{psiqu}) are closed.
In such cases, the effect beyond GP theories arises 
only through the quantity $w_3=-2\phi^2 q_V$. 
The Lagrangian densities $\tilde{\cal L}_5^{\rm N}$ 
and ${\cal L}_6^{\rm N}$, which are associated with the 
intrinsic vector modes, modify the quantity $q_V$. 
This modification affects the evolution of $\Psi,\Phi,$ and $\psi$
in a way similar to that in BP theories \cite{Geff}.
For the modes deep inside the Hubble radius, the rhs of 
Eq.~(\ref{deltaeq}) can be neglected relative to 
its lhs, such that 
\be
\ddot{\delta}+2H\dot{\delta}-4\pi G_{\rm eff} \rho_m 
\delta \simeq 0\,.
\label{mattersub}
\ee
Using the approximation $\delta \rho_m \simeq \rho_m \delta$ in 
Eq.~(\ref{Psiqu}), the effective gravitational coupling can be
estimated as 
\be
G_{\rm eff}=\frac{{\cal F}_1}{4\pi \phi \mu_5}\,.
\label{Geffquasi}
\ee
It is possible to rewrite $G_{\rm eff}$ by using physical quantities 
like $q_S$ and $c_{\rm S}^2$. In BGP theories with $\alpha_{\rm P}=0$, 
the form of $G_{\rm eff}$ is exactly the same as Eq.~(5.29) of 
Ref.~\cite{Geff}. Analogous to what happens in 
GP theories \cite{Geff}, there is a tendency that  
$G_{\rm eff}$ gets smaller for $q_V$ 
approaching $0^{+}$. 

{}From Eqs.~(\ref{Psiqu}) and (\ref{Phiqu}), the effective 
gravitational potential (\ref{Phieff}) under the quasistatic 
approximation reads 
\be
\Phi_{\rm eff}=\frac{{\cal F}_1+{\cal F}_2 \phi}
{\phi \mu_5}\frac{a^2}{k^2} \rho_m \delta\,.
\label{Phieffquasi}
\ee
In General Relativity, the quantity $({\cal F}_1+{\cal F}_2 \phi)/(\phi \mu_5)$ is 
equivalent to $8\pi G$, but the same quantity generally varies 
in time in BGP theories with $\alpha_{\rm P}=0$. Moreover, 
the different growth of $\delta$ affects the evolution of $\Phi_{\rm eff}$.
In BGP theories with $f_4 \neq 0$ and $f_5 \neq 0$, we have 
$\alpha_{\rm P} \neq 0$, so the terms containing 
$\alpha_{\rm P}$ in Eqs.~(\ref{mu5})--(\ref{F3}) 
lead to the modifications to $\Psi, \Phi, \psi$. 
In such cases, Eqs.~(\ref{Geffquasi}) and (\ref{Phieffquasi}) contain 
the time derivatives $\dot{\psi}$ and $\dot{\Phi}$. 
Solving the full perturbation equations to compute 
$\dot{\psi}, \dot{\Phi}$ and substituting them into 
Eqs.~(\ref{Psiqu})--(\ref{psiqu}), we can check 
whether the resulting values of $\Psi,\Phi,$ and $\psi$ 
reproduce those derived by the full numerical integration.
In Sec.\,\ref{growthconsec}, we will do so 
in the covariantized EVG model.

\subsection{Evolution of scalar perturbations 
in the covariantized EVG model} 
\label{growthconsec}
%

\begin{figure}
\begin{center}
\includegraphics[height=3.3in,width=3.3in]{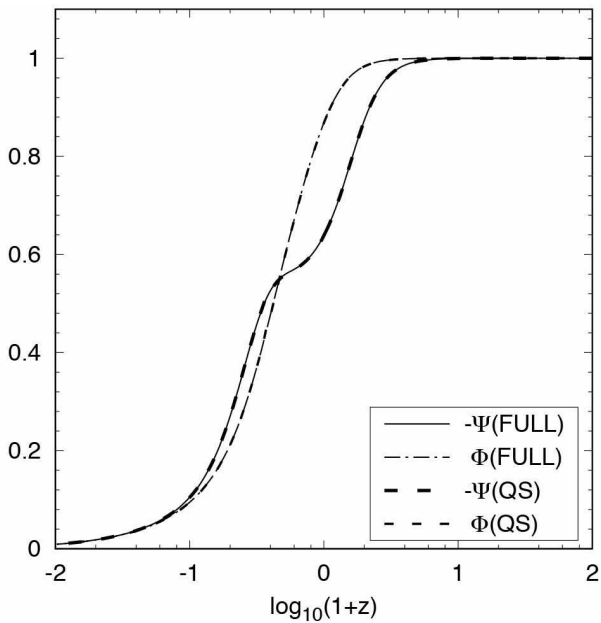}
\end{center}
\caption{\label{fig4}
Evolution of the gravitational potentials $-\Psi$ and $\Phi$ 
(normalized by their initial values) in covariantized and 
covariant EVG models for the mode $k=230a_0H_0$
with the model parameters 
$\beta_4=5.00 \times 10^{-2}, \beta_5=6.78 \times 10^{-2}$,  
$p_2=1$, $p=5$, $\lambda=1$, and 
the vanishing functions $g_5, G_6, \tilde{f}_5,$ and $\tilde{f}_6$. 
The initial conditions of perturbations are chosen to satisfy 
Eqs.~(\ref{Psiqu})--(\ref{psiqu}) with $\dot{\psi}=0$ and 
$\dot{\Phi}=0$ at $z=500$. Together with the full numerical solutions, we also show 
the results based on the quasistatic approximation (denoted as ``QS'' inside the figure) 
derived by substituting the full numerical 
solutions of $\dot{\psi}$ and $\dot{\Phi}$ into the rhs of Eqs.~(\ref{Psiqu}) and (\ref{Phiqu}).}
\end{figure}

In the covariantized EVG model, the quantity 
$\alpha_{\rm P}$ is given by 
\be
\alpha_{\rm P}=\frac{6[(1-2p-2p_2)\beta_4+(p+p_2)\beta_5]}
{(p+p_2)[1+6(1-2p-2p_2)\beta_4
+2(3p+2p_2)\beta_5]}\,,
\ee
which does not vanish for nonzero values of 
$\beta_4$ and $\beta_5$. 
Since $\alpha_{\rm P}$ is constant unlike
the quantity $\beta_{\rm P}$, the evolution of 
$\Psi, \Phi,$ and $\psi$ is affected by the presence of 
$\alpha_{\rm P}$-dependent terms in 
Eqs.~(\ref{Psiqu})--(\ref{psiqu}).

In Fig.~\ref{fig4}, we plot the full numerical solutions of $-\Psi$ and $\Phi$ in the 
covariantized EVG model with $k=230a_0H_0$ 
for $\beta_4=5.00 \times 10^{-2}, 
\beta_5=6.78 \times 10^{-2}$, $p_2=1$, $p=5$, $\lambda=1$, 
and vanishing values of $g_5, G_6, \tilde{f}_5,$ and $\tilde{f}_6$ 
(i.e., $q_V=1$). We choose the initial conditions satisfying
Eqs.~(\ref{Psiqu})--(\ref{psiqu}) with $\dot{\psi}=0$ 
and $\dot{\Phi}=0$. At high redshifts, the background matter 
density dominates over the vector-field density, 
so that the conditions (\ref{Acon}) are satisfied.
Provided that $|\beta_4| \ll 1$ and $|\beta_5| \ll 1$, the terms containing 
$\alpha_{\rm P}$ in scalar perturbation equations of motion 
are also suppressed in the early Universe relative to those associated with 
the background. Since ${\cal F}_1/(\phi \mu_5) \simeq {\cal F}_2/\mu_5 \simeq 4\pi G$ 
in this regime, the gravitational potentials in Eqs.~(\ref{Psiqu})--(\ref{Phiqu}) 
behave as $-\Psi \simeq \Phi \simeq 4\pi G 
(a^2/k^2) \rho_m \delta$ for $z \gg 1$. 
Since the matter density contrast evolves as $\delta \propto a$ 
during the deep matter era, we have that 
$-\Psi \simeq \Phi \simeq$\,constant in this regime. 

In the late Universe, the dynamics of $\Psi,\Phi,$ and $\psi$ is modified 
by the growth of the density of vector derivative interactions.
In Fig.~\ref{fig4}, we observe that the gravitational potentials 
$-\Psi$ and $\Phi$ start to vary at low redshifts 
with the parameter $\eta=-\Phi/\Psi$ 
deviating from 1. By solving the full perturbation equations 
numerically, we compute the time derivatives $\dot{\psi}$ and $\dot{\Phi}$
and substitute them into Eqs.~(\ref{Psiqu})--(\ref{Phiqu}).
As we see in Fig.~\ref{fig4}, the solutions derived under this 
approximation scheme exhibit good agreement with the 
full numerical results. We confirm that this is also the case for 
the matter perturbation equation (\ref{mattersub}) with the effective 
gravitational coupling (\ref{Geffquasi}).
If the terms $\epsilon_{\psi}$ and $\epsilon_{\Phi}$ are ignored 
in Eqs.~(\ref{Psiqu}), (\ref{Phiqu}), and (\ref{Geffquasi}), 
there are some deviations from 
the full numerical solutions 
at late times. Hence, the derivative terms $\dot{\psi}$ and $\dot{\Phi}$ 
should be included for deriving the solutions to 
the subhorizon perturbations accurately.
This means that the ``quasistatic'' approximation does not hold 
in the usual sense for the theories with $\alpha_{\rm P} \neq 0$.

Let us proceed to the discussion of the effective gravitational 
coupling $G_{\rm eff}$ and the matter density contrast $\delta$. 
In BGP theories with $f_4=0=f_5$, the BGP modifications to 
scalar perturbations arise only from $\tilde{f}_6$ 
through the quantity $w_3=-2\phi^2 q_V$. 
In the covariantized EVG model with Eq.~(\ref{G2F}), 
the quantity $q_V$ is given by $q_V=1+4\tilde{f}_6H^2 \phi^2$.
On using the expression of $G_{\rm eff}$ given in Eq.~(5.29) of 
Ref.~\cite{Geff}, it is possible to realize $G_{\rm eff}<G$ for 
$0<q_V \ll 1$ in BGP theories with $\alpha_{\rm P}=0$.
For the function $\tilde{f}_6=c_6X^{q_6}$, the quantity 
$q_V$ reduces to Eq.~(\ref{qVexa}), so $q_V$ can be 
a positive constant for $q_6=p-1$. 
As shown in Ref.~\cite{Geff}, however, the realization of 
$G_{\rm eff}$ smaller than $G$ requires that $q_V$ is 
quite close to zero.
Moreover, the deviation of $G_{\rm eff}$ from 
$G$ is not so significant that it is still difficult for it to 
be compatible with the RSD data (see the left panel of Fig.~2 of Ref.~\cite{Geff}). 
We can also consider the time-varying functions $q_V$ 
(say, $q_6>p-1$ and $c_6<0$), but in such cases, we require 
further tunings to ensure the stability condition $q_V>0$.

\begin{figure*}
\begin{center}
\includegraphics[height=3.2in,width=3.3in]{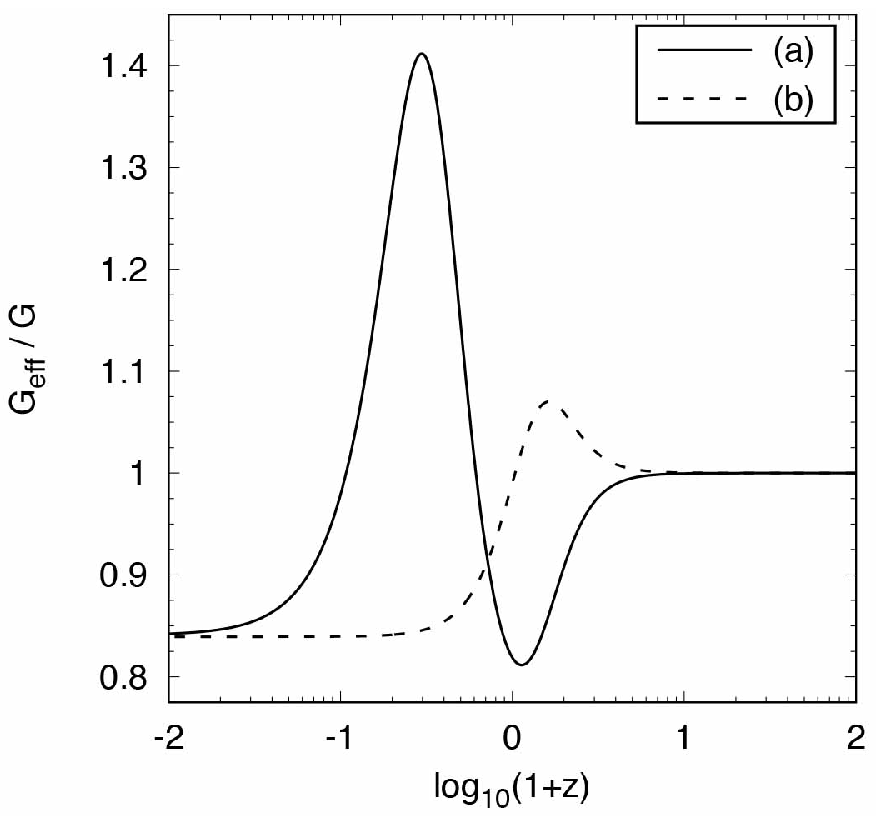}
\includegraphics[height=3.2in,width=3.3in]{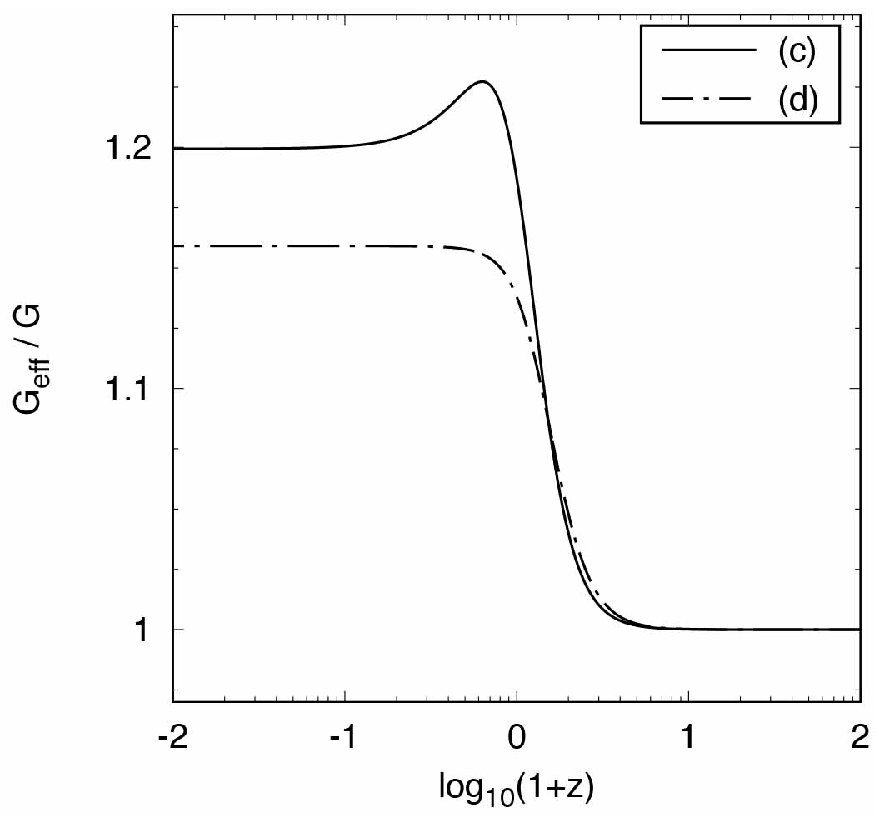}
\end{center}
\caption{\label{fig5}
(Left) Evolution of $G_{\rm eff}/G$ for the wave number 
$k=230a_0H_0$ with the model parameters 
$\beta_4=5.00 \times 10^{-2}$, $\beta_5=6.78 \times 10^{-2}$, 
$p_2=1$, $p=5$, $\lambda=1$, and the vanishing functions 
$g_5, G_6, \tilde{f}_5,$ and $\tilde{f}_6$. 
The initial conditions of perturbations are chosen to satisfy 
Eqs.~(\ref{Psiqu})--(\ref{psiqu}) with $\dot{\psi}=0$ and 
$\dot{\Phi}=0$ at $z=500$.
Cases (a) and (b) correspond to the covariantized and  
covariant EVG models, respectively. In case (a), the weak 
gravity ($G_{\rm eff}<G$) is realized by today, while in 
case (b), $G_{\rm eff}$ temporally increases and finally 
approaches a value smaller than $G$ on the de Sitter attractor.
(Right) Evolution of $G_{\rm eff}/G$ 
for $\beta_4=1.00 \times 10^{-2}$ and $\beta_5=0$ 
with the other parameters the same as those used in the left panel. 
Cases  (c) and (d) are the covariantized and 
covariant EVG models, respectively, both of which 
correspond to the strong gravity ($G_{\rm eff}>G$).
In case (c), the asymptotic value of $G_{\rm eff}/G$ 
on the de Sitter solution is different from that in case (d).
}
\end{figure*}

In BGP theories with nonvanishing functions $f_4$ and $f_5$, 
the additional terms arising from $\alpha_{\rm P}$ to 
scalar perturbation equations of motion can modify the 
evolution of $G_{\rm eff}$ at low redshifts. 
To understand the effect of the $\alpha_{\rm P}$ term, 
we first consider the covariant EVG model 
and then discuss the covariantized EVG model later.
In GP theories, the value of $G_{\rm eff}$ on the de Sitter solution 
is generally given by \cite{Geff}
\be
(G_{\rm eff})_{\rm dS}=\frac{H(2H\phi^2q_V-w_6\phi-w_2)}
{4\pi [(2H\phi^2 q_V-w_6 \phi)(w_2+2Hq_T)
+w_1w_2]}\,.
\label{Geffes}
\ee
In the covariant EVG model with $p_2=1, p=5$ and 
$g_5=0=G_6$ (i.e., $q_V=1$), 
for example, Eq.~(\ref{Geffes}) reduces to 
\begin{widetext}
\be
(G_{\rm eff})_{\rm dS}
=G\frac{(1-11\beta_4+4\beta_5)[2-108\beta_4+56\beta_5 
+(1-11\beta_4+4\beta_5)(u_{\rm dS})^2]}
{2+1584\beta_4^2+8\beta_5(8+37\beta_5)
-12\beta_4 (11+116\beta_5)+6(11\beta_4-6\beta_5) 
(1-11\beta_4+4\beta_5)(u_{\rm dS})^2}\,.
\label{GeffdSes}
\ee
\end{widetext}

\begin{figure}
\begin{center}
\includegraphics[height=3.1in,width=3.3in]{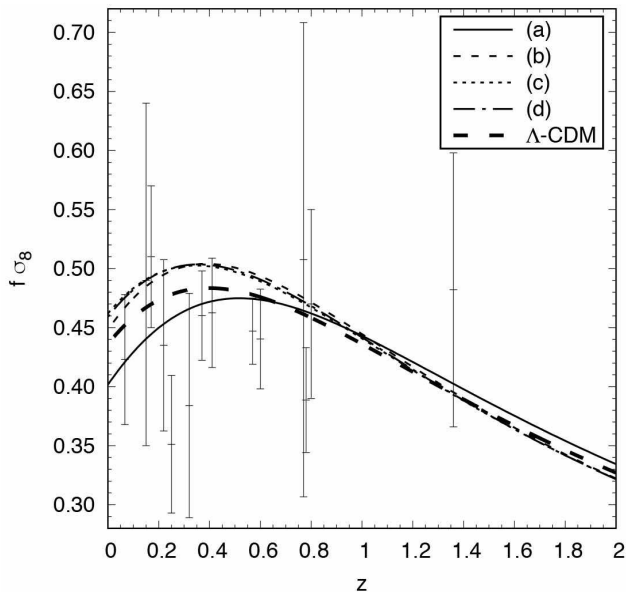}
\end{center}
\caption{\label{fig6}
Evolution of $f\sigma_8$ vs the redshift $z$ (in the regime 
$0 \le z \le 2$) for the four cases (a), (b), (c), and (d) corresponding 
to the models in Fig.~\ref{fig5}. 
The initial conditions of perturbations are chosen to 
satisfy Eqs.~(\ref{Psiqu})--(\ref{psiqu}) and 
$\dot{\psi}=0, \dot{\Phi}=0$
with the comoving wave number 
$k = 230a_0H_0$ and $\sigma_8(z=0)=0.82$. 
The evolution of $f\sigma_8$ in the $\Lambda$CDM model 
is plotted as a dashed bold line.
We also show the bounds of $f\sigma_8$ with error bars 
constrained from the RSD measurements \cite{2dF,Guzzo,6dF,Wiggle,SDSS,BOSS,Torre,Okumura}. 
}
\end{figure}

Case (b) shown in Fig.~\ref{fig5} corresponds to 
the covariant EVG model with
$\beta_4=5.00 \times 10^{-2}$, $\beta_5=6.78 \times 10^{-2}$, 
and $u_{\rm dS}=1.193$, so that 
$(G_{\rm eff})_{\rm dS}=0.839G$ from Eq.~(\ref{GeffdSes}).
Although $G_{\rm eff}<G$ on the de Sitter attractor, 
$G_{\rm eff}$ temporally grows from the value close to $G$ 
after the end of the matter era, and then it starts to decrease toward 
the value smaller than $G$.
Since $G_{\rm eff}>G$ during most of the epoch by today, 
the growth rate of $\delta$ in case (b) is larger than 
that in the $\Lambda$CDM model for $z \geq 0$.
This property can be confirmed by the numerical integration of 
$f\sigma_8$ plotted in Fig.~\ref{fig6}, 
where $f \equiv \dot{\delta}/(H\delta)$ and 
$\sigma_8$ is the amplitude of $\delta$ at the comoving 
$8\,h^{-1}$ Mpc scale ($h$ is the normalized Hubble 
constant $H_0 = 100 h$ km sec$^{-1}$\,Mpc$^{-1}$). 
The values of $f\sigma_8$ in case (b) 
are larger than those of the $\Lambda$CDM model 
in the redshift range $0 \leq z \lesssim 1$.

In BGP theories with $\alpha_{\rm P} \neq 0$, 
the estimation (\ref{Geffes}) loses its validity. 
The time derivatives $\dot{\psi}$ and 
$\dot{\Phi}$ generally approach zero toward the de Sitter solution, 
but the same property also holds for $\psi$ and $\Phi$. 
Although the two quantities $\epsilon_{\psi}$ and $\epsilon_{\Phi}$ 
should be finite on the de Sitter attractor, their 
values are not known $a~priori$.

In case (a) of Fig.~\ref{fig5}, we plot the evolution of 
$G_{\rm eff}/G$ in the covariantized EVG model 
with $q_V=1$ for the same parameters $\beta_4,\beta_5,p_2,$ and $p$ 
as those used in case (b). 
While the values of $G_{\rm eff}$ on the de Sitter solution 
are similar to each other between cases (a) and (b), 
the significant difference arises during the transition from the end of the 
matter era to the de Sitter attractor. 
In case (a), $G_{\rm eff}$ first decreases to reach a minimum 
with $G_{\rm eff} \simeq 0.8\,G$ at the redshift around $z=0$. 
After the temporal increase of $G_{\rm eff}$ toward the regime 
$G_{\rm eff}>G$ in the future, the effective gravitational coupling 
finally approaches a value smaller than $G$.
Unlike case (b), the weak gravity ($G_{\rm eff}<G$) 
can be realized by today.

As we see in case (a) of Fig.~\ref{fig6}, the values of 
$f\sigma_8$ at low redshifts are smaller than those of 
the $\Lambda$CDM model. 
By using the best-fit value of $\sigma_8(z=0)$ 
constrained by the Planck CMB measurement \cite{Planckdark},  
case (a) 
can be compatible with most of the recent RSD data.
This behavior arises from the existence of nonvanishing 
terms $\alpha_{\rm P}$ beyond the domain of GP theories.
Thus, the BGP theories offer an interesting possibility of realizing 
weak gravitational interactions consistent with the 
RSD measurements.

The evolution of $G_{\rm eff}$ is subject to modifications 
for different choices of $\beta_4$ and $\beta_5$.
In the right panel of Fig.~\ref{fig5}, we plot the 
evolution of $G_{\rm eff}/G$ in the covariantized EVG 
model [case (c)] and in the covariant EVG model [case (d)]
for $\beta_4=1.00 \times 10^{-2}$ and $\beta_5=0$ 
with the other model parameters the same as those used in the left panel.
Since $u_{\rm dS}=1.172$ in these cases, the estimation 
(\ref{GeffdSes}) gives $(G_{\rm eff})_{\rm dS}=1.159G$ 
on the de Sitter solution in the covariant EVG model. 
In case (d) of Fig.~\ref{fig5}, $G_{\rm eff}$ starts to evolve 
from the value close to $G$ and then it continuously grows 
to the asymptotic value $1.159G$. 
In case (c), the existence of nonvanishing terms $\alpha_{\rm P}$ 
leads to a different value of $G_{\rm eff}\,(\simeq 1.2G)$ on the de Sitter solution. 
As we see in Fig.~\ref{fig5}, the effective gravitational 
coupling in case (c) is also larger than $G$ during the 
cosmic expansion history.
Since the growths of $G_{\rm eff}/G$ in cases 
(c) and (d) are similar to each other for $z \geq 0$, the values of 
$f\sigma_8$ are also degenerate. 
In Fig.~\ref{fig6}, cases (c) and (d) do not fit the RSD data 
very well due to the property $G_{\rm eff}>G$.

\begin{figure}
\begin{center}
\includegraphics[height=3.2in,width=3.3in]{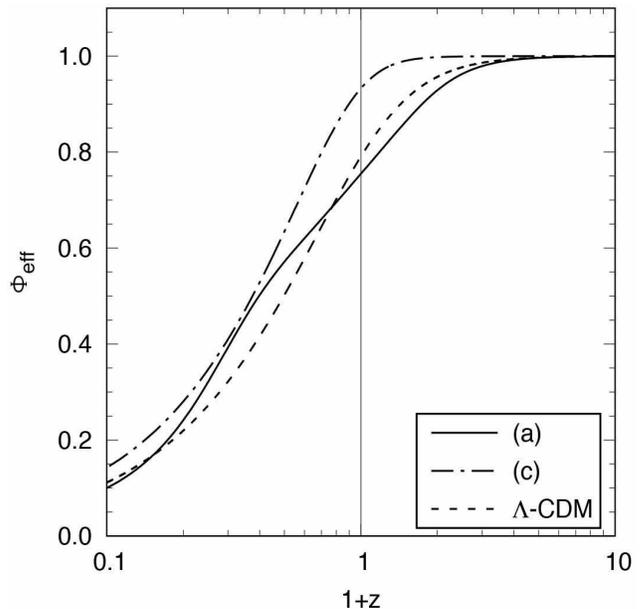}
\end{center}
\caption{\label{fig7}
Evolution of the weak lensing gravitational potential 
$\Phi_{\rm eff}$ (normalized by its initial value) 
for cases (a) and (c) shown 
in Fig.~\ref{fig5} and for the $\Lambda$CDM model. 
The present epoch ($z=0$) is shown as a vertical thin line.}
\end{figure}

In Fig.~\ref{fig7}, we plot the evolution of $\Phi_{\rm eff}$ defined by 
(\ref{Phieff}) in the covariantized EVG model for cases (a) and (c) 
in Fig.~\ref{fig5}. 
The weak lensing gravitational potential in case (a) decreases faster 
than that in the $\Lambda$CDM model for $z\geq 0$, 
whereas in case (c), $\Phi_{\rm eff}$ initially exhibits 
tiny growth and starts to decrease by today.
This difference arises from the different evolution of
$\Psi$ as well as $\eta$.
We expect that future observations of weak lensing offer the 
possibility of distinguishing between the covariantized EVG model and 
the $\Lambda$CDM model.

\section{Conclusions}
\label{consec}

We have studied the cosmology in BGP theories with five 
propagating degrees of freedom  (one scalar, two vectors, and 
two tensors) on the flat FLRW background. 
Compared to second-order GP theories with the Lagrangian 
densities (\ref{L2})--(\ref{L6}), there are four additional derivative interactions 
given by Eqs.~(\ref{L4N})--(\ref{L6N}).
The latter interactions are detuned to keep 
the equations of motion up to second order, but 
they still do not cause the Ostrogradski instability 
with the Hamiltonian unbounded from below.

At the background level, the equations of motion 
(\ref{back1})--(\ref{back3}) contain four functions 
$A_{2,3,4,5}$ defined by Eq.~(\ref{ABre}). 
In GP theories, they are associated with the four functions 
$G_{2,3,4,5}$ in ${\cal L}_{2,3,4,5}$.
In BGP theories, the additional functions 
$f_4$ and $f_5$, which are related to the intrinsic 
scalar mode, also arise
from ${\cal L}_4^{\rm N}$ and  ${\cal L}_5^{\rm N}$.
Introducing the functions $B_4$ and $B_5$ as 
Eq.~(\ref{ABre}), there are two relations 
(\ref{ABrelation0}) and (\ref{ABrelation}) 
between $A_4, A_5, f_4,$ and $f_5$. 
Since $f_4=0=f_5$ in GP theories, the functions 
$B_4$ and $B_5$ are directly related to $A_4$ and $A_5$. 
In BGP theories, the existence of two free functions 
$B_4$ and $B_5$ leads to modifications to the evolution 
of cosmological perturbations. 
Moreover, the additional two functions
$\tilde{f}_5$ and $\tilde{f}_6$ in $\tilde{\cal L}_5^{\rm N}$ 
and ${\cal L}_6^{\rm N}$, 
which are associated with intrinsic vector modes, also affect the dynamics of 
vector and scalar perturbations.

Since our interest is the application of BGP theories to the 
late-time cosmic acceleration, we have explored the cosmological 
dynamics for a concrete dark energy 
scenario called the covariantized EVG model. 
In GP theories, there is a counterpart dubbed the covariant 
EVG model. In these two models, the functions $A_{2,3,4,5}$ 
are the same, but the functions $B_{4,5}$ are different, 
i.e., Eq.~(\ref{extendedco2}) for the covariant EVG and 
Eq.~(\ref{extendedco3}) for the covariantized EVG. 
Hence, the background expansion history is the same in 
both cases with the dark energy equation of 
state given by Eq.~(\ref{wde}).
Since the background solution is characterized by the 
phantom equation of state during the matter era 
($w_{\rm DE}=-1-s$ with $s=p_2/p>0$) followed by a de Sitter 
attractor, these two models can be clearly distinguished 
from the $\Lambda$CDM model.

In Sec.\,\ref{tensec}, we discussed theoretically consistent 
conditions of tensor perturbations in the covariantized EVG model.
While the no-ghost condition is the same as that in the covariant 
EVG model, the stability condition is different due to a modification 
of the tensor propagation speed. 
Provided that the normalized constants $\beta_4$ and $\beta_5$ 
defined by Eq.~(\ref{betai}) are in the range (\ref{beta45}), 
there are neither ghosts nor Laplacian instabilities in 
both covariantized and covariant EVG models.

In Sec.\,\ref{vecsec}, we studied no-ghost and stability 
conditions of vector perturbations in the small-scale limit. 
The intrinsic vector modes arising from 
$\tilde{{\cal L}}_5^{\rm N}$ and ${\cal L}_6^{\rm N}$ 
lead to modifications to the quantities $q_V$ and $c_V^2$ 
relative to those in GP theories.
As long as the conditions (\ref{f6con}) 
and (\ref{f5con}) are satisfied for $\tilde{f}_6<0$ 
and $\tilde{f}_5>0$, it is possible to avoid the appearance of 
ghosts and Laplacian instabilities in BGP theories with (\ref{G2F}).

In Sec.\,\ref{scasec}, we derived no-ghost and stability conditions 
of scalar perturbations in the presence of radiation and 
nonrelativistic matter.
In BGP theories, the scalar propagation speed $c_{\rm S}$ arising 
from the longitudinal mode of the vector field is coupled to 
the matter sound speeds, the mixing of which is weighed by the 
parameter $\beta_{\rm P}$. 
The quantity $\beta_{\rm P}$ is proportional to the combination 
$f_4+3H\phi f_5$, which vanishes in GP theories ($f_4=0=f_5$).
We studied the evolution of $c_{\rm S}^2$ from the radiation era 
to the de Sitter epoch in the covariantized EVG model and showed 
that the mixing is suppressed in such a way that 
$c_{\rm S}^2$ is practically equivalent to the decoupled value 
$c_{\rm P}^2$ with $|c_{\rm P}^2| \gg |\beta_{\rm P}|$.
Analytically, we obtained the values of $c_{\rm S}^2$ 
during radiation, deep matter, and de Sitter epochs 
and derived the stability conditions (\ref{stacon1})--(\ref{stacon3}) 
in the limit that $|\beta_4| \ll 1, |\beta_5| \ll 1$. 
We also found that the difference of $c_{\rm S}^2$ between 
covariantized and covariant EVG models mostly comes from the 
different choices of the functions $B_4$ and $B_5$ in $c_{\rm P}^2$.

In Sec.\,\ref{effsec}, we investigated the evolution of matter density 
contrast and gravitational potentials for the subhorizon 
perturbations associated with the observations of large-scale 
structures and weak lensing.
On using the so-called quasistatic approximation, we showed 
that the existence of BGP Lagrangian densities 
${\cal L}_4^{\rm N}$ and ${\cal L}_5^{\rm N}$ gives 
rise to time derivatives $\dot{\psi}$ and 
$\dot{\Phi}$, while they do not appear in GP theories.
Hence, the perturbation equations for the scalar degree of 
freedom $\psi$ and gravitational potentials $\Psi$ and $\Phi$ are 
not closed even under this approximation scheme. 
In BGP theories, we need to solve the full perturbation equations of motion 
in order to know the evolution of perturbations accurately. 
Computing the time derivatives $\dot{\psi}$ and $\dot{\Phi}$ by the full 
numerical integration and substituting them 
into Eqs.~(\ref{Psiqu}) and (\ref{Phiqu}), they can reproduce 
the full numerical solutions to $\Psi$ and $\Phi$; see Fig.~\ref{fig4}. 

In both covariantized and covariant EVG models, we studied 
the evolution of the effective gravitational coupling 
$G_{\rm eff}$ and the growth rate of matter perturbations. 
Even when the values of $G_{\rm eff}$ on the de Sitter 
attractor are similar to each other between the two models, 
the behavior of $G_{\rm eff}$ during the transition from the matter era 
to the de Sitter epoch is generally different
(e.g., the left panel of Fig.~\ref{fig5}).
In the covariantized EVG model, it is possible to realize the 
situation in which $G_{\rm eff}$ decreases to the value like 
$G_{\rm eff} \simeq 0.8\,G$ by today. 
In this case, the growth rate of matter perturbations is 
smaller than that in the $\Lambda$CDM model, 
so the covariantized EVG model can be compatible 
with the recent RSD data of 
$f\sigma_8$ even by using the Planck best fit of 
$\sigma_8(z=0)$; see Fig.~\ref{fig6}.
This behavior of weak gravity occurs
by the existence of the BGP derivative interactions
${\cal L}_4^{\rm N}$ and ${\cal L}_5^{\rm N}$. 

In the covariant EVG model, the existence of intrinsic 
vector modes allows the possibility of $G_{\rm eff}<G$, 
but this requires that the quantity $q_V$ is quite 
close to zero \cite{Geff}. Moreover, the values of $G_{\rm eff}$ 
in the redshift range $0 \le z<1$ are not significantly smaller than 
$G$ in general, so the realization of weak gravity in the covariant 
EVG model is limited compared to the 
covariantized EVG model. Hence, it is possible to distinguish 
between the two models from the $f\sigma_8$ data of
RSD measurements. 
Depending on the model parameters, the covariantized EVG model 
can also lead to $G_{\rm eff}$ larger than $G$ (like 
the right panel of  Fig.~\ref{fig5}), so it may be possible 
to exclude some parameter spaces from the RSD data.
The weak lensing gravitational potential $\Phi_{\rm eff}$
also exhibits the difference from that in the 
$\Lambda$CDM (see Fig.~\ref{fig7}), 
so this information can be used to place constraints
 on the covariantized EVG model further.

We have thus shown that BGP theories allow the construction of 
a concrete dark energy model with the equation 
of state $w_{\rm DE}$ smaller than $-1$, 
while the growth rate of 
matter perturbations can be compatible with the RSD 
data by reflecting the property $G_{\rm eff}<G$. 
A similar attempt was carried out in GLPV 
scalar-tensor theories \cite{weakgra}, 
but it was later found that the model proposed for realizing $G_{\rm eff}<G$ 
is plagued by the problem of solid-angle-deficit singularities at the center 
of a spherically symmetric body \cite{conical}. 
In BGP theories, solid-angle-deficit singularities do not 
generally arise due to the existence of a temporal vector component \cite{HKTconi}. 
It remains to be seen whether future 
high-precision observations including RSD and weak lensing 
show some evidence that the covariantized EVG model is 
favored over the $\Lambda$CDM model.

\section*{Acknowledgements}

We thank Lavinia Heisenberg for useful comments.
R. K. is supported by the Grant-in-Aid for Research Activity
Start-up of the JSPS (Grant No.\,15H06635). 
S. T.  is supported by the Grant-in-Aid for Scientific Research Fund of the JSPS (Grant No.~16K05359) and 
MEXT KAKENHI Grant-in-Aid for 
Scientific Research on Innovative Areas ``Cosmic Acceleration'' (Grant No.\,15H05890).

\appendix
\section{SCALAR PROPAGATION SPEED FOR THE COVARIANT EVG MODEL}

In this Appendix, we compute the scalar propagation 
speed squared $c_{\rm S}^2$ for the covariant EVG model 
in the limits $\Omega_{\rm DE} \to 0$ (radiation and 
early matter eras) and 
$\Omega_{\rm DE} \to 1, \Omega_r \to 0$ 
(de Sitter era). Since $\beta_{\rm P}$ vanishes in this case, 
$c_{\rm S}^2$ is exactly equivalent to $c_{\rm P}^2$.
During the radiation, early matter, and de Sitter eras, 
we obtain the following values of $c_{\rm S}^2$, respectively:
\begin{widetext}
\ba
\hspace{-0.8cm}
(c_{\rm S})_r^2
&=& \frac{2-3 p-4 p_{2}-2 \beta_{5} (3 p+2 p_{2}-1) (3 p+4 p_{2}-6) + 6 \beta_{4} [6+6 p^{2}+8 (p_{2}-2) p_{2}+p (14 p_{2}-17)]}{3 p^{2} [6 (2 p+2 p_{2}-1) \beta_{4} -(4 p_{2} + 6  p) \beta_{5}-1]} \,,\label{csr_covariant} \\
\hspace{-0.8cm}
(c_{\rm S})_m^2
&=& \frac{3-5 p-6 p_{2}-2 \beta_{5} [9 + 3 p (5 p - 11 ) +4 p_{2} (3p_{2}+7 p-6)]+6 \beta_{4} [9+ p(10 p-27 )    +2 p_{2} (6p_{2}+11 p-12)]}{6 p^{2} [6 (2 p+2 p_{2}-1) \beta_{4} -(4 p_{2} + 6  p) \beta_{5}-1]},
\label{csm_covariant}\\
\hspace{-0.8cm}
(c_{\rm S})_{\rm dS}^2
&=&\frac{ \xi [(p + p_{2}) \xi -  \gamma \{\gamma + (1 + p) (1 - 2 p_{2} \beta_{5})\} (q_{V} u^{2})_{\rm dS}  ]}{6 \gamma^{2} (2 p_{2} \beta_{5} - 1) \{\gamma+p  (1-2  p_{2} \beta_{5})\} 
(q_{V} u^{2})_{\rm dS} }
\label{csds_covariant} \,,
\ea
where
\ba
\xi
~& &\equiv
p_{2} [\gamma+(1+p) (1-2 p_{2} \beta_{5})] [1+6 (5-2 p-2 p_{2}) \beta_{4}-2 (6-3 p-2 p_{2}) \beta_{5}]
\nonumber \\
& &~~~+ [ \gamma+p  (1-2 p_{2} \beta_{5})]
[ \gamma+ (p-1) (1-2 p_{2} \beta_{5})].
\ea
\end{widetext}


\end{document}